\documentclass[12pt]{amsart}
\usepackage{amssymb}
\usepackage{amsmath}
\usepackage{graphicx}
\usepackage{soul}
\usepackage{color}
\begin{document}

\title[Fast cycles detecting]{Fast cycles detecting in non-linear discrete systems}

\author{D.Dmitrishin, E.Franzheva and A.Stokolos}

\begin{abstract}
In the paper below we consider a problem of stabilization of a priori unknown unstable periodic orbits in non-linear autonomous discrete dynamical systems. We suggest a generalization of a non-linear DFC scheme to improve the rate of detecting $T$-cycles. Some numerical simulations are presented.
\end{abstract}
\maketitle

\section{Introduction}

Chaotic regimes are typical for many non-linear dynamical systems that simulate processes in various areas of research, e.g. in physics, economics, ecology, electronics, etc. (c.f. \cite{CD}). A characterization of such regimes is the existence of infinitely many unstable periodic orbits. Stabilization of such orbits is one of the important tasks in
non-linear control theory \cite{FE}.\\

If a periodic orbit is known then the most popular method of stabilization is the OGY method \cite{OGY}. To stabilize a priori unknown periodic orbits various delay feedback control schemes have been developed (c.f. \cite{P, VL, M1, DK, DKKS, DKST}). Certain limitations of DFC schemes were mentioned in, e.g., \cite{U}. \\

One of the methods that allows to stabilize at least in theory an unstable orbit is a generalization of the method of non-linear feedback (NDFC) by Viera-Licheberg \cite{VL}. Its deficiencies are related to a narrow basin of attraction for stabilized periodic solutions as well as  with slow rates of convergence of perturbed solutions to the periodic one. In  the paper below we suggest a possible modification of the NDFC scheme \cite{DK, DKKS, DKST} that allows one to increase the rate of convergence. The gain coefficients are chosen such that cycle multipliers are contained in the central disc of radius $\rho<1.$  Thus, the rate of convergence will be of order at least $\rho^n, n\to \infty.$\\

It is clear that we have to pay a price for that. Namely, in construction we use a non-linear DFC or mixing with the length of prehistory increasing with the decreasing of the size of the region of the associated multiplier's location. Moreover, if $\rho=1$ then for any size of region there exists delay that allows to use in control the prehistory of certain length to stabilize the cycle. However, if $\rho<1$ then that is not a case. For a given size of the diameter of the multiplier's region there exists a limited value $\rho_0$ that does not guaranties the  convergence at a rate of order smaller then $\rho_0^n, n\to\infty$ regardless of the length of the used prehistory.\\

In other words, for  a given dynamical system with $\rho$ decreasing the length of the prehistory used in the control increases to infinity when $\rho\to \rho_0.$ However, the effectiveness of modified control is very well visible on the examples provided in the paper.\\

\section{Settings and preliminary results}

Let us consider the vector non-linear discrete dynamical system
\begin{equation} \label{dsc}
x_{n+1}=F(x_n),\qquad  x_n\in\mathbb R^m,\quad n=1,2,...
\end{equation}
It is assumed that the system \eqref{dsc} has invariant convex sets $A,$ i.e. if $\xi\in A$ then $F(\xi)\in A.$ It is also assumed that this system has an unstable $T-$ cycle $\left(\eta_{1},\, \, \ldots \, , \eta_{T}\right)$, where all vectors $\eta_{1},\, \, \ldots \, ,  \eta_{T}$ are pairwise distinct and belongs to the invariant set $A,$ i.e.
$\eta_{j+1}=F(\eta_{j}), j=1,\ldots,T-1, \eta_{1}=F(\eta_{T})$.\\

The multipliers  $\{\mu_1,..,\mu_m\}$ considered  for the cycle are determined as eigenvalues  of a product of Jacobi matrices $\prod_{j=1}^T F'(\eta_j).$  The $T$ cycle is asymptotically stable if and only if all multipliers are in the open unit disc of the complex plane.\\

We  are interested in detecting cycles of arbitrary length. Let us note that even in a simplest case of a scalar polynomial function $F$ the detecting of cycle of the length $T$ by $T$ self-iterations of the function $F$ does not work if $T$ is large. Indeed, the $T$-iterated function will be a polynomial of order the initial degree risen to the power $T$. The second problem can be extraneous cycles obtained that way. We have to increase the depth of the prehistory. It is impossible to keep the length of prehistory one as is the case with Viera-Lightenebrg or  Pyragas controls.
\iffalse %%%%%%%%%%%
   Since in applications the most used functions $F$ are polynomials then to detect a cycle of length $T$ one should solve the polynomial equation of degree  $N^T,$ where $N$ is a degree of the polynomial $F$ which is a serious obstacle if $T$ is large. Moreover, even detection of one point in the cycle does not guarantee that all other points can be found by direct substitution of the value in the system \eqref{dsc} because of rounding error and potential strong instability of the cycle.
   \fi%%%%%%%%%%%%%%%
   \\

To avoid the above obstacles we suggest  construct a new system that has same cycles but such that they are stable. We are looking for a system either in the form of
\begin{equation}\label{2}
x_{n+1}=\sum_{i=1}^N a_i f(x_{n-iT+T}),
  \end{equation}
or of the form
\begin{equation}\label{2.1}
x_{n+1}=f\left(\sum_{i=1}^N a_i x_{n-iT+T}\right),
  \end{equation}
where $a_1+...+a_N=1.$\\

Let us note that the system \eqref{2} is obtained from the system $$
x_{n+1}=f(x_n)+u_n,
$$
where $u_n$ is control based on the non-linear feedback with delay, i.e.
$$
u_n=-\sum_{j=1}^{N-1}\epsilon_j\left(f(x_{n-jT+T})-f(x_{n-jT})\right)
$$
while the control in the system \eqref{2.1} is organized on the mixing principle. c.f. \cite{DSS}.\\

So, our goal is to make locally stable the $T-$ cycles of the system \eqref{2} and \eqref{2.1}. It is important that the convex set $A$ is still invariant for the systems
\eqref{2} and \eqref{2.1} as well.
% , i.e. if vectors $\xi_0,\xi_T,\xi_{2T},\ldots,\xi_{(N-1)T}$ are contained in set $A,$ then the vector $\sum_{j=1}^M\gamma_j f(\sum_{i=1}^N \alpha_{ij}\xi_{(N-i)T})$ is contained in the set $A$ too.
This follows from the definition of the convex combination of the vectors. On a top of that the systems \eqref{2} and \eqref{2.1} have same $T$-cycles that are in the system \eqref{dsc}.\\

The characteristic equation for the cycle of above systems \eqref{2} and \eqref{2.1} is found in \cite{DHKS}
\begin{equation} \label{che}
\prod_{j=1}^m\left( \lambda^{(N-1)T+1}-\mu_j(a_1\lambda^{N-1}+...+a_N)^T \right)  =0.
\end{equation}
The stability condition is that the roots of the characteristic equation \eqref{che}
lie in the unit disc. Thus it is required to find a number $N$ and  coefficients $(a_1,...,a_N)$ such that all polynomials of the one-parametric family
\begin{equation} \label{sch}
\left\{ \lambda^{(N-1)T+1}-\mu(a_1\lambda^{N-1}+...+a_N)^T: \mu\in M \right\}
\end{equation}
are Schur stable. Here $M$ is a set of location of multipliers.\\%, i.e. we study {\it robust stability}. \\

Therefore, we come up with the following problem: {\it for given cycle length $T$ and given set of multipliers localization define the coefficients of  mixing $a_{i},i=1,\ldots,N$  such that cycle of length $T$ will be locally asymptotically stable; the magnitude of using prehistory being be minimum possible.}\\

Clearly, the solution of problem depends on the localization of the set of multipliers  $\{\mu_1,\ldots,\mu_m\}$.\\

We will consider two possibilities: either all multipliers are real negative
$$\{\mu_1,\ldots,\mu_m\}\subset\{\mu\in \mathbb R:\mu\in(-\mu^*,0)\},\; \mu^*>1,$$
or  are complex and located in the left half-plane
$$\{\mu_1,\ldots,\mu_m\}\subset\{\mu\in \mathbb C:|\mu+R|<R\},\; R>1/2.$$

%For each of these cases the algorithm of finding minimum $N$ and optimal coefficients $\{a_1,\ldots,a_N\}$ consists of the following steps \cite{18}:

Finding optimal values for the coefficients $a_1,...,a_N$ turns out to be a difficult problem. It is completely solved for $T=1,2$ for real multipliers \cite{DK, DKKS} and $T=1$ in case of complex multipliers with negative real part \cite{DHKS}. There is a strong numeric evidence that the coefficients suggested in \cite{DKST} are optimal for all $T$ and the above multipliers. They are defined in the following way.

\begin{itemize}
                \item [a)] compute nodes:
                $$
                \psi_j=\frac{{\pi(\sigma+T(2j-1))}}{\sigma+(N-1)T}, j=1,2,\ldots,\frac{N-2}{2},\;\mbox{if $N$ is even};\; \frac{N-1}{2}, \mbox{if $N$ is  odd};
                $$
                In the case $\{\mu_1,\ldots,\mu_m\}\subset\{\mu \in\mathbb R:\mu\in(-\mu^*,0)\}$ we pick $\sigma=2$, while in the case $\{\mu_1,\ldots,\mu_m\}\subset\{\mu\in \mathbb C:|\mu+R|<R\}$  we pick $\sigma=1$;
                \item[b)] construct auxiliary polynomials
                $$
                \eta_N(z)=z(z+1)\prod_{j=1}^{\frac{N-2}{2}}(z-e^{i\psi_j})(z-e^{-i\psi_j}), \;\mbox{$N$ even},
                $$
                $$
                \eta_N(z)=z\prod_{j=1}^{\frac{N-1}{2}}(z-e^{i\psi_j})(z-e^{-i\psi_j}),\;\mbox{ $N$ odd;}
                $$
                \item[c)] compute coefficients of auxiliary polynomials
                $$
                \eta_N(z)=\sum_{j=1}^{N}c_jz^j;
                $$
                \item[d)] construct {\it standard} coefficients
                \begin{equation}\label{st}
                a_j=\frac{(1-\frac{1+(j-1)T}{2+(N-1)T})c_j}{\sum_{k=1}^{N}(1-\frac{1+(k-1)T}{2+(N-1)T})c_k},\; j=1,\ldots,N;
                \end{equation}
                \item[e)] in case $\{\mu_1,\ldots,\mu_m\}\subset\{\mu\in \mathbb R:\mu\in(-\mu^*,0)\}$ compute values
                $$
                I_N^{(T)}=-\left[\frac{T}{2+(N-1)T}\prod_{k=1}^{\frac{N-2}{2}}\cot^2\frac{\pi(2+T(2k-1))}{2(2+(N-1)T)}\right]^{T},\;\mbox{$N$ even}
                $$
                $$
                I_N^{(T)}=-\left[\prod_{k=1}^{\frac{N-1}{2}}\cot^2\frac{\pi(2+T(2k-1))}{2(2+(N-1)T)}\right]^{T},\; \mbox{$N$ odd};
                $$
                The optimal value of $N$ is computed as minimal positive integer that satisfies the inequality
                \begin{equation} \label{opt1}
                \mu^*\cdot |I_N^{(T)}|<1;
                \end{equation}
                \item[f)] in case $\{\mu_1,\ldots,\mu_m\}\subset\{\mu\in \mathbb C:|\mu+R|<R\}$ compute values
                $$
                I_N^{(T)}=-\left[\frac{T}{1+(N-1)T}\prod_{k=1}^{\frac{N-2}{2}}\cot^2\frac{\pi(1+T(2k-1))}{2(1+(N-1)T)}\right]^{T},\;\mbox{$N$ even},
                $$
                $$
                I_N^{(T)}=-\left[\prod_{k=1}^{\frac{N-1}{2}}\cot^2\frac{\pi(1+T(2k-1))}{2(1+(N-1)T)}\right]^{T},\;\mbox{$N$ odd;}
                $$
                The optimal value of $N$ is computed as minimal positive integer that satisfies the inequality
\begin{equation} \label{opt2}
                R\cdot 2|I_N^{(T)}|<1.
\end{equation}
\end{itemize}

It was found in \cite{DK, DKKS} that in case of real multipliers
\begin{equation} \label{opt12r}
                |I_N^{(1)}|=\tan^2\frac\pi{2(N+1)},\qquad
                |I_N^{(2)}|=\frac1{N^2}.
\end{equation}
In other words the minimal value $N$ should satisfy the inequality
\begin{equation} \label{optmur}
\mu^*<\cot^2\frac\pi{2(N+1)}\quad\mbox{or}\quad \mu^*<N^2,
\end{equation}
while in the case of complex multipliers
\begin{equation} \label{opt1c}
                |I_N^{(1)}|=\frac1{N}
\end{equation}
and the minimal value $N$ should satisfy the inequality
\begin{equation} \label{optmuc}
2R<N.\end{equation}

For example, in the case e) above the {\it standard} coefficients $(a_1,...,a_N)$ can be chosen as the following
\begin{equation}\label{aj1}
a_j=2\tan\frac\pi{2(N+1)}\left(1-\frac j{N+1}\right)\sin\frac{\pi j}{N+1},\quad j=1,...,N
\end{equation}
for the case $T=1$ and as the following
\begin{equation}\label{aj2}
a_j=\frac2{N}\left(1-\frac {2j-1}{2N}\right)\,\quad j=1,...,N
\end{equation}
for the case $T=2.$ \\

In the case f) above the {\it standard} coefficients are the following
\begin{equation}\label{aj3}
a_j=\frac2{N}\left(1-\frac {j}{N+1}\right)\,\quad j=1,...,N.
\end{equation}
We will call those coefficients optimal. Let us stress one more time that the coefficients defined by the formula \eqref{st} in both real and complex cases are called {\it {\it standard}}.\\

They are optimal in the sense of widest range for multipliers,
but they might be far from optimal in the sense of rate of convergence. Indeed, in the case of real multipliers and $T=1$ if the value of $\mu^*$ is close to $\cot^2\frac\pi{2(N+1)}$ then the roots of the polynomial $f(\lambda)=\lambda^N+\mu^*(a_1\lambda^{N-1}+...+a_N)$ might be close to the boundary of the unit disc, therefore the convergence of the iterative procedures \eqref{2} and \eqref{2.1} will be quite slow.\\

Furthermore, if the roots are in the disc of radius $\rho\le1$ the rate of convergence might be slow because the {\it standard} coefficients are designed to serve the worse case scenario. This situation is very well illustrated on the pictures below in the section 4.

A natural question emerges - how to increase the rate of convergence? Let us post a problem: find a positive integer $N$ and the coefficients $(b_1,...,b_N)$ such that all polynomials of the one-parametric family \eqref{sch} have roots inside the disc of radius
$\rho<1.$\\

\section{Fast stabilization result}

Let us consider the polynomials of the family  \eqref{sch}. If  the coefficients $a_1,...,a_N$ coincide with the {\it {\it standard} } then all roots of these polynomials are inside the unit disc $\mathbb D.$  At that case the value $N$ determining the length of prehistory is minimal. Let us demand that the roots of the considered polynomials be inside the disc of radius $\rho<1.$ It is clear that the value $N$ will be more than the {\it standard}. Let us find the corresponding coefficients $b_1,...,b_N$ which we will call {\it modified}.

First, let us consider the case of real multipliers and $T=1.$
To solve this problem let us make a substitution $\lambda=z\rho.$ It is clear that $|\lambda|<\rho$ if and only if $|z|<1.$ Denote $p(\lambda)=a_1\lambda^{N-1}+...+a_N.$ The equation $\lambda^N-\mu p(\lambda)=0$ is equivalent to the equation
$$
z^N-\mu\frac{p(\rho)}{\rho^N}\frac{p(z\rho)}{p(\rho)}=0,
$$
or
\begin{equation}\label{phat}
z^N-\hat\mu\hat p(z)=0,
\end{equation}
where $\hat\mu=\mu\frac{p(\rho)}{\rho^N}$ and $\hat p(z)=\frac{p(z\rho)}{p(\rho)}.$\\

The roots of the polynomial \eqref{phat} have to be in the unit disc. By \eqref{optmur} this happens if  $\hat\mu^*<\cot^2\frac\pi{2(N+1)}$ and $\hat p(z)=a_1^{(N)}z^{N-1}+...+a_N^{(N)},$ where $a_1^{(N)},...,a_N^{(N)}$ are {\it standard} coefficients. Since $\hat p(z)=\frac{p(z\rho)}{p(\rho)}$ then
$$
\frac{p\left(\frac1\rho \rho\right)}{p(\rho)}=\hat p\left(\frac1\rho\right).
$$
Since $p(1)=1,$ then $\frac1{p(\rho)}=\hat p\left(\frac1\rho\right).$\\

Further
$$
\mu^*=\hat\mu^*\frac{\rho^N}{p(\rho)}=\hat\mu^*\rho^N
\hat p\left(\frac1\rho\right)<\rho \hat q(\rho)\cot^2\frac\pi{2(N+1)},
$$
where $z\hat q(z)=z^N\hat p(\frac1z),$ i.e. $\hat q(z)=a_1^{(N)}+...+a_N^{(N)}z^{N-1}.$\\

Further, $\frac{p(\lambda)}{p(\rho)}=\hat p\left(\frac\lambda\rho\right),$ from there
$$
p(\lambda)=p(\rho)\hat p\left(\frac\lambda\rho\right)=
\frac1{\hat p\left(\frac1\rho\right)}\hat p\left(\frac\lambda\rho\right).
$$
Therefore, the solution is the following:

i. The minimal value of $N$ should satisfy the inequality
$$
\mu^*<(a_1^{(N)}\rho+...+a_N^{(N)}\rho^N)\cot^2\frac\pi{2(N+1)},
$$
where the coefficients $(a_1^{(N)},...,a_N^{(N)})$ are {\it standard}, i.e. determined by \eqref{aj1};\\

ii. The optimal polynomial is
$$
p_O(\lambda)=\frac1{\hat p\left(\frac1\rho\right)}
\hat p\left(\frac\lambda\rho\right)=
\frac{\rho^N}{\rho\hat q(\rho)}\hat p\left(\frac\lambda\rho\right)=
b^{(N)}_1\lambda^{N-1}+...+b^{(N)}_N.
$$
where the optimal modified coefficients are
$$
b^{(N)}_j=\frac{a_j^{(N)}\rho^j}{\sum_{k=1}^N a_k^{(N)}\rho^k}.
$$
\\

The same approach allows us to consider $T$ cycles in real and in complex case, where $\mu$ might belong to $(-\mu^*,0)$ as well as $|\mu+R|<R$         too. Namely, the equation $\lambda^{(N-1)T+1}-\mu [p(\lambda)]^T=0$ is
equivalent to the following
$$
z^{(N-1)T+1}-\mu\frac{[p(\rho)]^T}{\rho^{(N-1)T+1}}
\left[\frac{p(z\rho)}{p(\rho)}\right]^T=0
$$
or
\begin{equation}\label{cheT}
z^{(N-1)T+1}-\hat\mu [\hat p(z)]^T=0,
\end{equation}
where $\lambda=z\rho,\hat\mu=\mu\frac{[p(\rho)]^T}{\rho^{(N-1)T+1}}$ and $\hat p(z)=\frac{p(z\rho)}{p(\rho)}.$\\

The roots of the equation \eqref{cheT} have to be in the central unit disc which means that $\hat\mu^*<\frac1{|I^{(N)}_T|}$ in the real case and that
$\hat\mu^*\in\left\{ z\in\mathbb C: \left|z+\frac1{2|I^{(N)}_T|}\right|<\frac1{2|I^{(N)}_T|}\right\}$ in the complex case, and
$p(z)=a_1^{(N)}z^{N-1}+...+a_N^{(N)},$ where $a_j^{(N)}$ are {\it standard} coefficients for the general case. Since $\hat p(z)=\frac{p(z\rho)}{p(\rho)}$ then $\frac1{p(\rho)}=\hat p\left(\frac1\rho \right).$
%where$\hat p(z)=a_1^{(N)}z^{N-1}+...+a_N^{(N)}.$\\

Then
$$
\mu^*<\frac{\rho\left[\hat q(\rho)\right]^T}{|I^{(T)}_N|},
$$
or
$$
\mu\in\left\{z\in\mathbb C: \left|z+\frac{\rho\left[ \hat q(\rho)\right]^T}{2
|I^{(T)}_N|} \right|< \frac{\rho\left[\hat q(\rho)\right]^T}{2
|I^{(T)}_N|} \right\},
$$
where $\hat q(z)=a_1^{(N)}+...+a_N^{(N)}z^{N-1}$ and $z\hat q(z)$ is inverse to $\hat p(z).$ \\

Further,
$$
p(\lambda)=\frac1{\hat p\left(\frac1\rho\right)}
\hat p\left(\frac\lambda\rho\right)=\frac{\rho^N}{\rho\hat q(\rho)}\hat p\left(\frac\lambda\rho\right)=
b^{(N)}_1\lambda^{N-1}+...+b^{(N)}_N,
$$
where the optimal modified coefficients are
$$
b^{(N)}_j=\frac{a_j^{(N)}\rho^j}{\sum_{k=1}^N a_k^{(N)}\rho^k}.
$$

\section{Roots visualization for the standard and for the modified polynomials}

In this section we consider the problems of location of zeros of {\it standard} polynomial
$\hat f(\lambda)=\lambda^{(N-1)T+1}-\hat\mu \hat p(\lambda)$ and {\it modified} polynomials $ f_O(\lambda)=\lambda^{(N-1)T+1}-\mu \hat p_O(\lambda)$ for various $T$ and coefficients - the {\it standard} $(a_1^{(N)},...,a_N^{(N)} )$ and the  {\it modified} ones
$(b_1^{(N)},...,b_N^{(N)})$ computed in the case of real multipliers as well as complex.\\

\subsection{Homothecy property}
The conditions of stability of polynomials $\hat f(\lambda)$ and $ f_O(\lambda)$ in the space of parameters $\hat\mu$ and $\mu$ correspondingly are given by inclusions \cite{DKST}
$$
\hat\mu\in \left\{\left(\bar{\mathbb C}\backslash z\left(\sum_{j=1}^Na_j^{(N)}z^{j-1}
\right)^T\right)^*,z\in\mathbb D\right\},
$$
$$
\mu\in \left\{\left(\bar{\mathbb C}\backslash z\left(\sum_{j=1}^Nb_j^{(N)}z^{j-1}
\right)^T\right)^*,z\in\mathbb D\right\},
$$
where $\mathbb D$ denotes the unit disc and $*$ denotes inversion, i.e. $z^*=\frac1{\bar z},$ $\bar{\mathbb C}=\mathbb C \cup \{\infty\}.$\\

Let us show the examples of the inverse images of the central unit circle under the polynomial maps $z\left(\sum_{j=1}^Na_j^{(N)}z^{j-1}\right)^T$ and $z\left(\sum_{j=1}^Nb_j^{(N)}z^{j-1}\right)^T.$

\begin{center}
               \includegraphics[width=5cm]{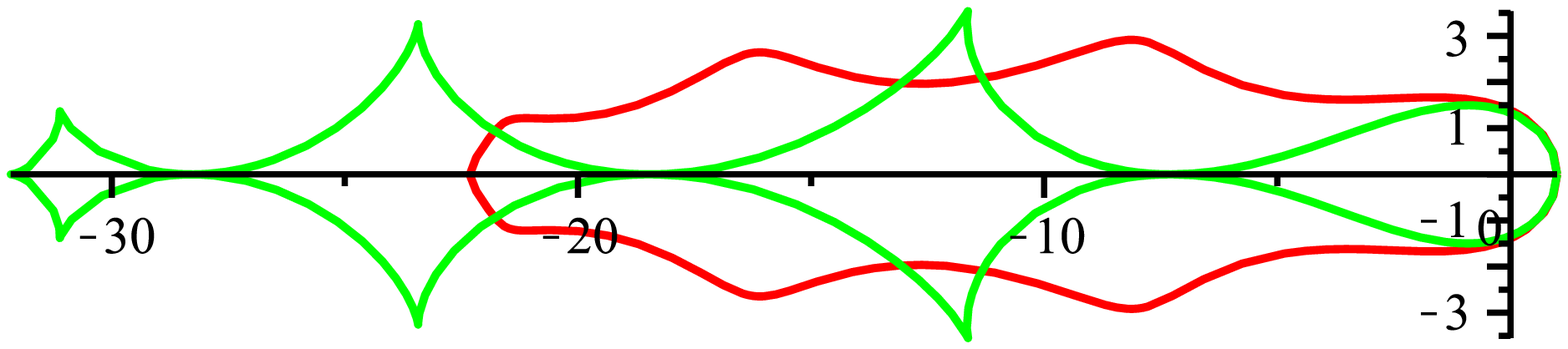}\hspace{2cm}
              \includegraphics[width=5cm]{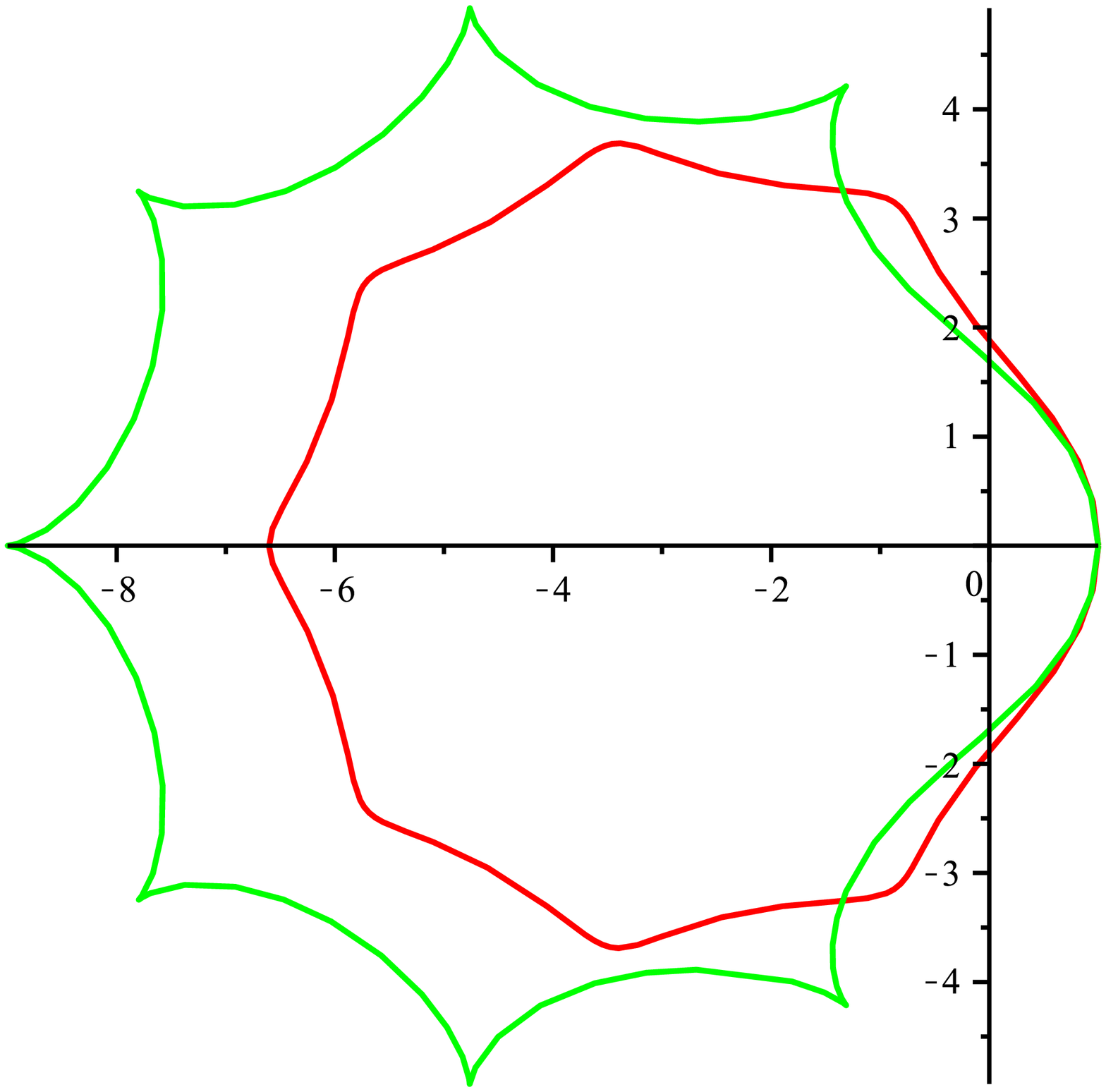}

   \hspace{-1cm}             Fig.1
\end{center}
The Fig.1 displays the inverse image of the unit disc under the {\it standard} map (green) and under the {\it modified} map (red) for $T=1, N=8,\rho=0.9$ The multipliers are real on the left figure and  complex on the right figure.

\begin{center}
               \includegraphics[width=5cm]{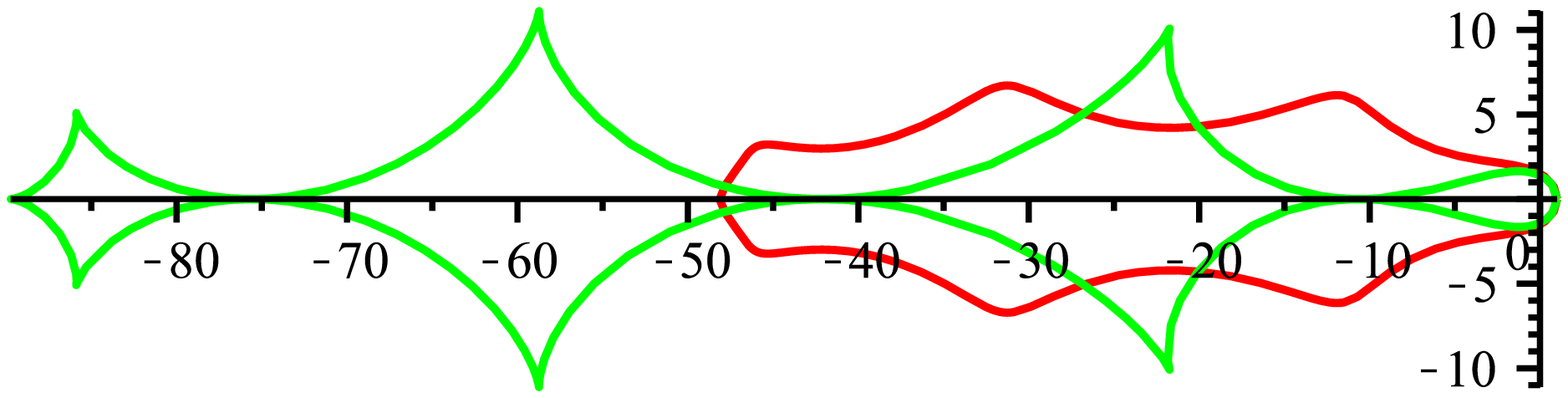}\hspace{2cm}
             \includegraphics[width=5cm]{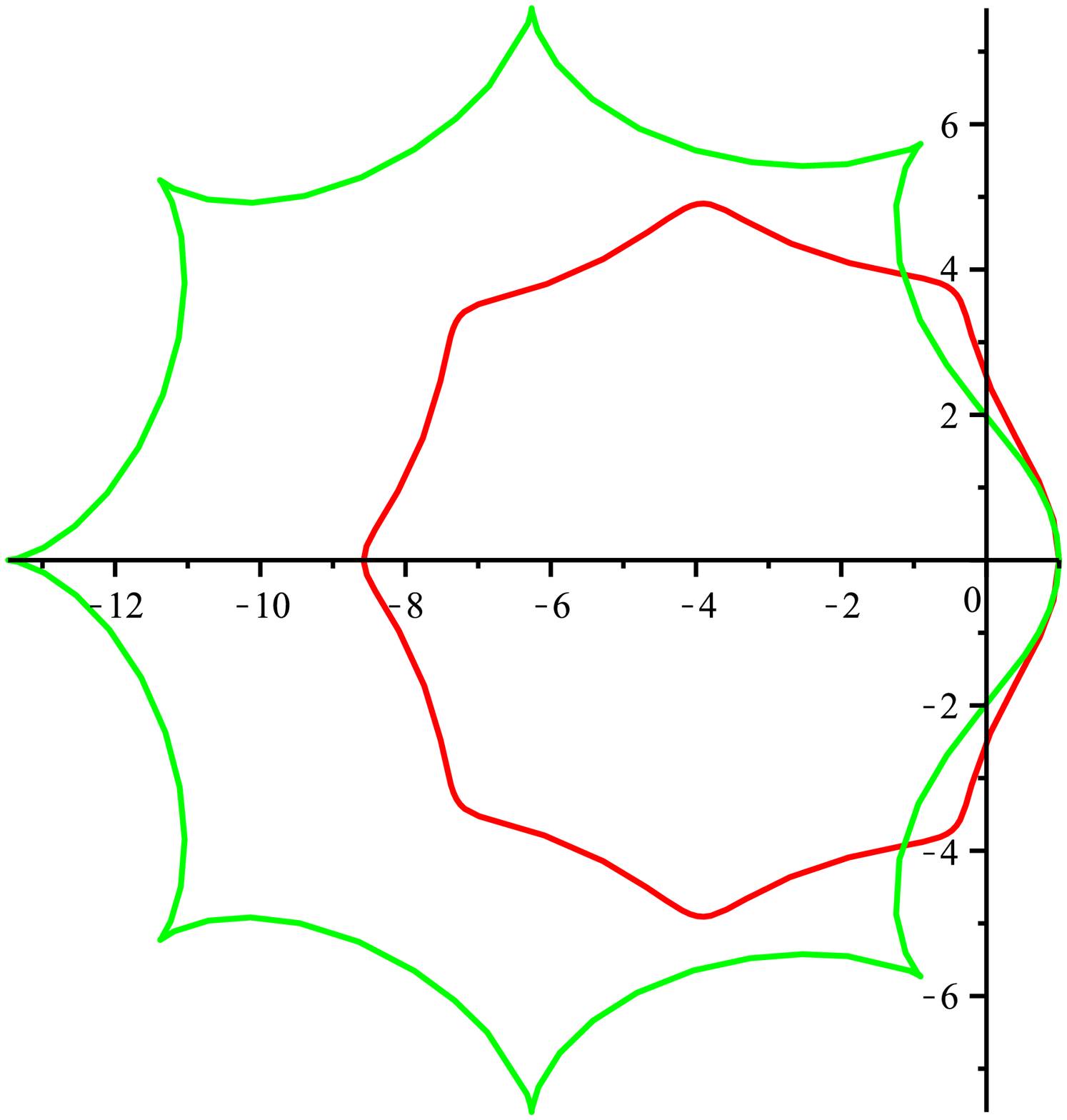}

   \hspace{-1cm}             Fig.2
\end{center}
The Figure 2 illustrates the inverse image of the unit disc under the {\it standard} map (green) and under the {\it modified} map (red) for $T=3, N=8,\rho=0.9$ The multipliers are real on the left figure and complex on the right figure.

Let us note that the following identity is valid
$$
\frac1{e^{it}}{\left(\sum_{j=1}^Na_j^{(N)}e^{i(j-1)t}\right)^T}=
\frac1{\rho\left(\sum_{j=1}^Na_j^{(N)}\rho^{j-1}\right)^T}
\frac1{\frac1\rho e^{it}\left(\sum_{j=1}^Nb_j^{(N)} \left(\frac1\rho e^{it}\right)^{j-1}\right)^T}.
$$
That means that the inverse image of the unit disc under the {\it standard} polynomial map is homothetic to the inverse image of the disc radius $\frac1\rho$ under the {\it modified} polynomial map with the coefficient of homothecy
$$
\frac1{\rho\left(\sum_{j=1}^Na_j^{(N)}\rho^{j-1}\right)^T}=
\frac1{\rho\left(\hat q(\rho)\right)^T}.
$$
Let us illustrate this phenomenon on examples.

\begin{center}
               \includegraphics[width=5cm]{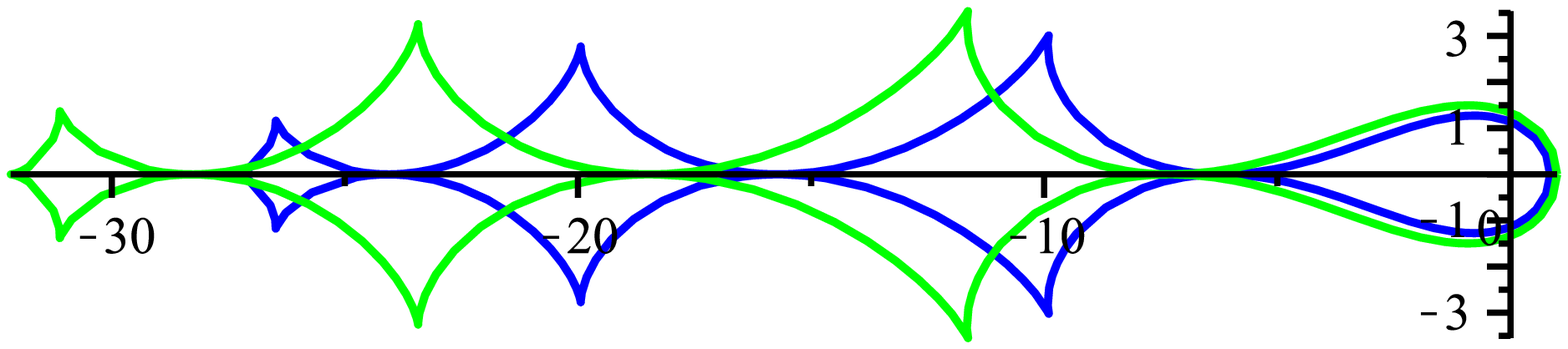}\hspace{2cm}
              \includegraphics[width=5cm]{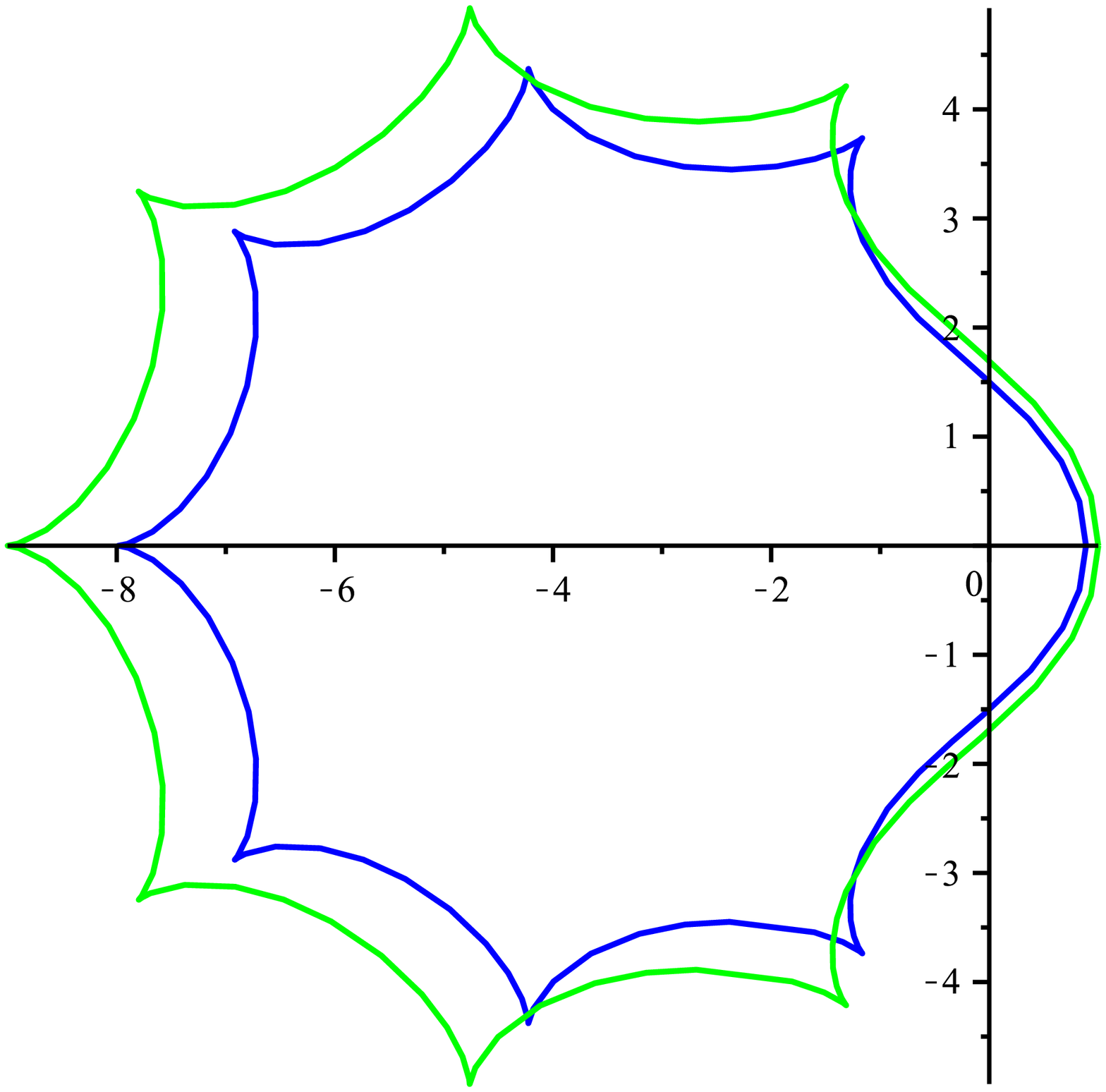}

   \hspace{-1cm}             Fig.3
\end{center}

The Figure 3 illustrates the inverse image of the unit disc under the {\it standard} map (green) and the inverse image of the disc of radius $\frac1\rho$ under the {\it modified} map (blue) for $T=1, N=8,\rho=0.9$ The multipliers are real on the left figure and are complex on the right figure.

\begin{center}
               \includegraphics[width=5cm]{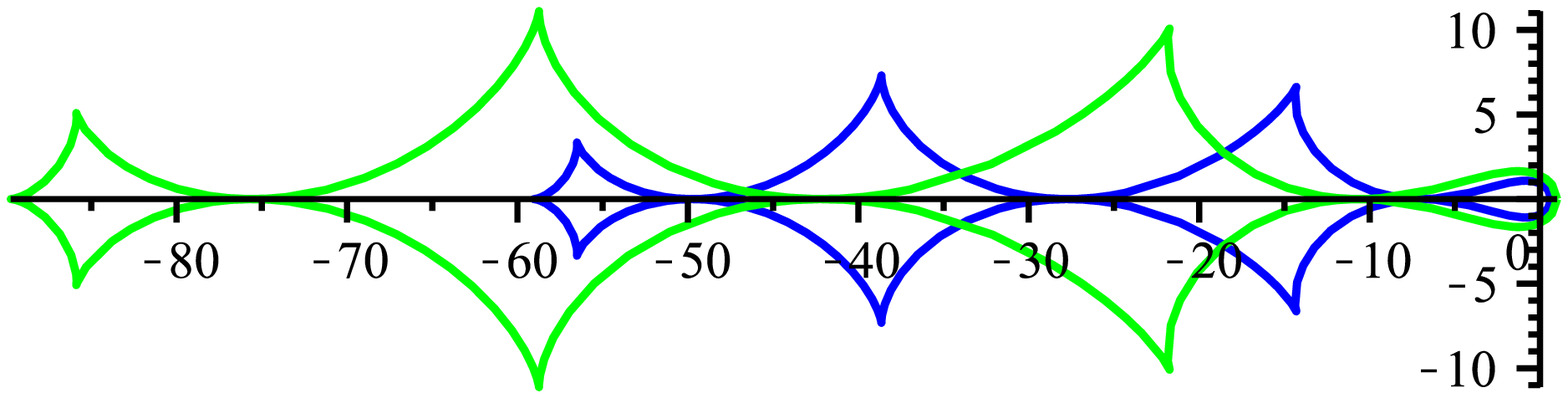}\hspace{2cm}
              \includegraphics[width=5cm]{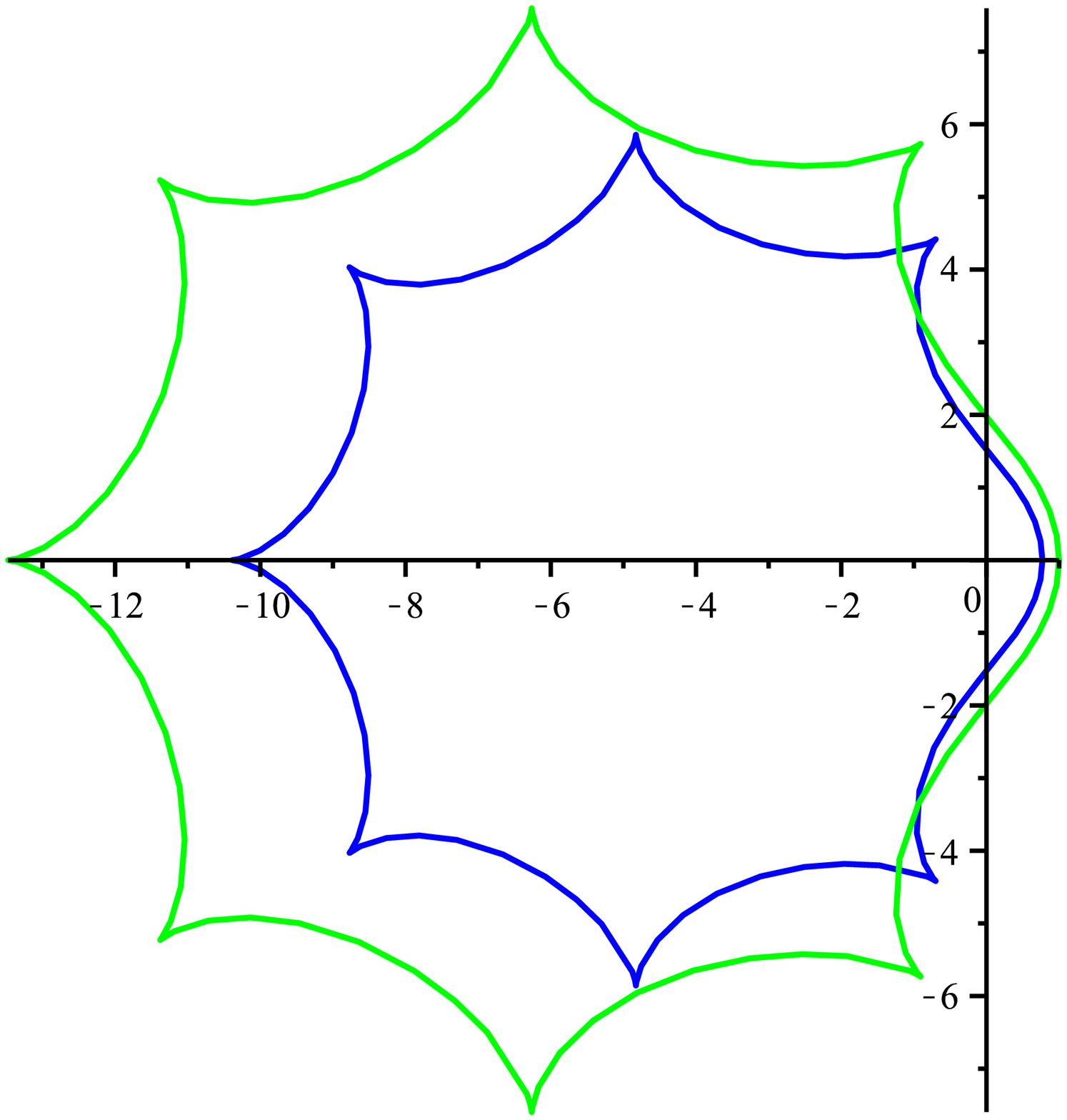}

   \hspace{-1cm}             Fig.4
\end{center}

The Figure 4 illustrates the inverse image of the unit disc under the {\it standard} map (green) and the inverse image of the disc of radius $\frac1\rho$ under the {\it modified} map (blue) for $T=3, N=8,\rho=0.9.$ The multipliers are real on the left figure and complex on the right figure.\\

\subsection{Root location. Standard polynomials.}
First, let us note that in particular case $m=1$ (scalar case) the degree of the equation \eqref{che} is $1+(N-1)T,$ e.g. if $N=8, T=1$ it is 8, if $N=8, T=2$ it is 15 and if   $N=8, T=3$ it is 22.\\

It is very well visible that for small values of $\mu$ the roots are far from the boundary. On the Figure 5-28 the value $N$ equals $8$, therefore one can observe 8, 15 and 22 differently colored zeros corresponding the case $T=1,2,3$ and different multipliers.\\

In the case of real multipliers and $T=1, N=8$ the critical value $\mu^*=\cot^2\frac\pi{18}=32.16343764...$ (see \eqref{optmur}). Thus, the roots  corresponding to the choice $\mu=-2$ (red), $\mu=-10$ (green), $\mu=-20$ (blue) are separated from the unit circle on Fig.5 while $\mu=-31$ (brown) is close to the critical value and the corresponding roots are close to the boundary, as it can be very well seen on Fig.6. \\

\begin{center}
               \includegraphics[width=4cm]{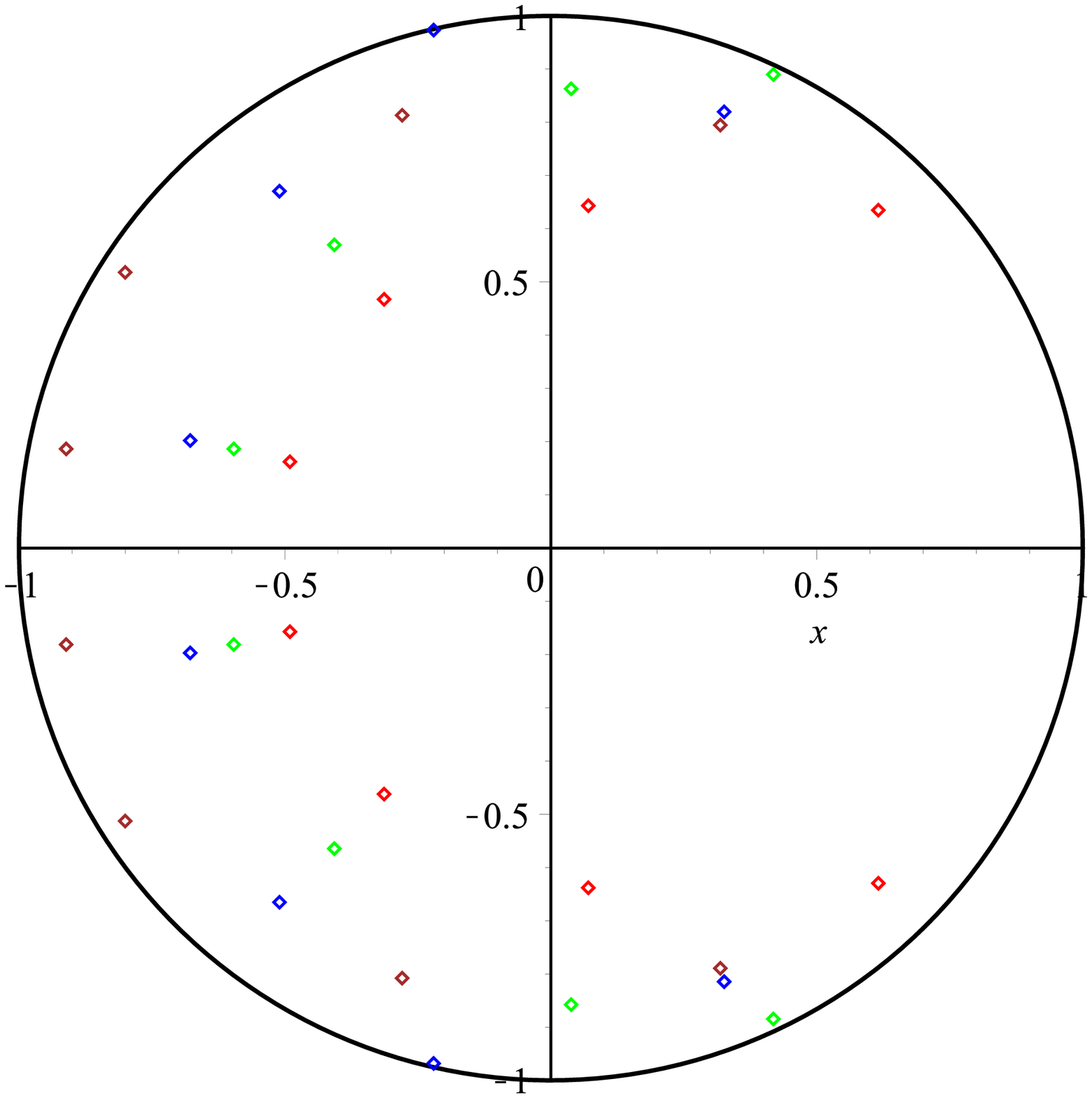}\hspace{2cm}
               \includegraphics[width=4cm]{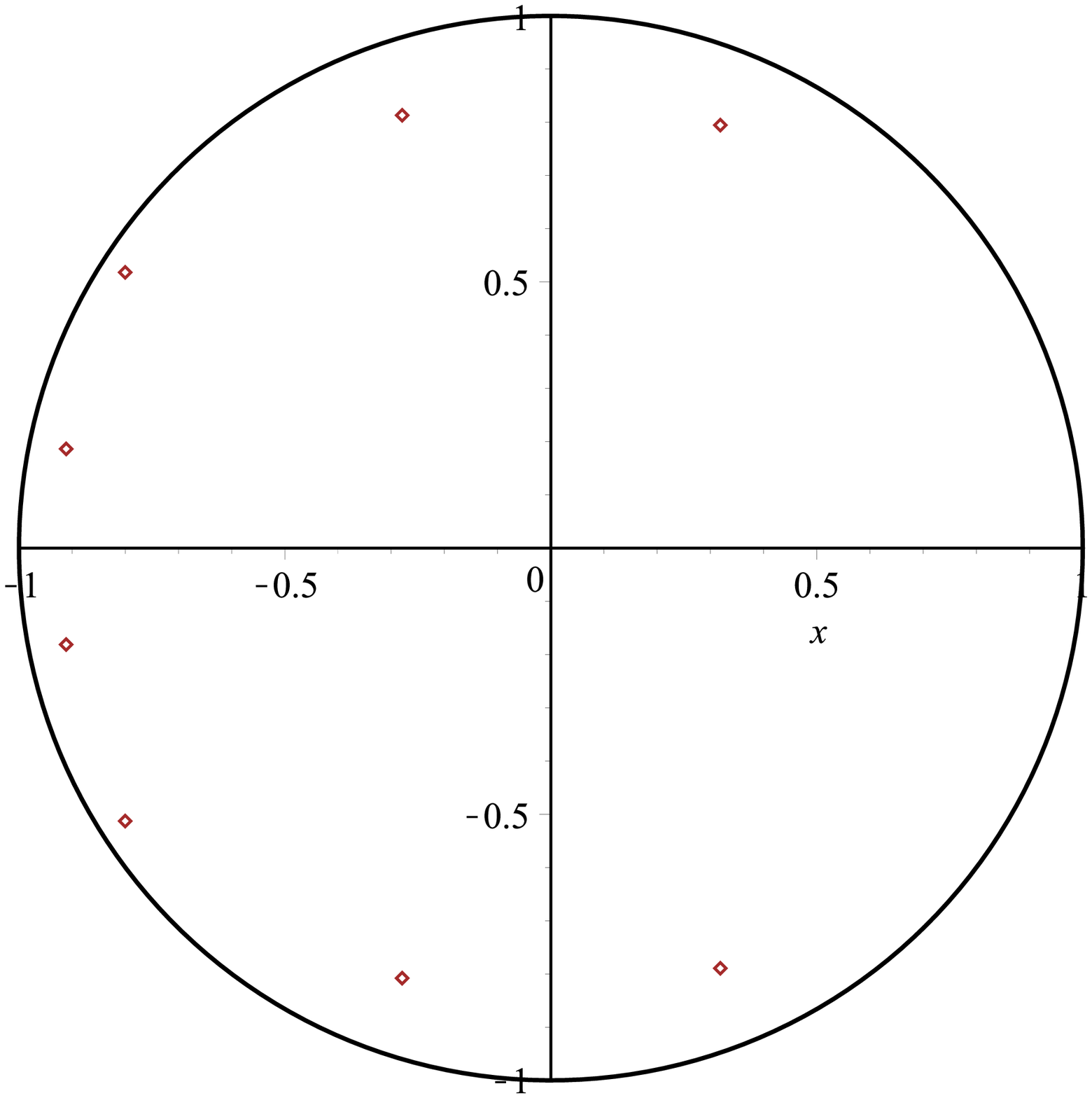}

   \hspace{-1cm}             Fig.5 T=1, $\mu=-2,-10,-20,-31$. Fig.6 T=1, $\mu=-31$
\end{center}

In the case of real multipliers and $T=2, N=8$ the critical value $\mu^*=8^2=64$ (see \eqref{optmur}). Thus, the roots  corresponding to the choice $\mu=-2$ (red), $\mu=-20$ (green), $\mu=-40$ (blue) are separated from the unit circle on Fig.7 while $\mu=-63$ (brown) is close to the critical value and the corresponding roots are close to the boundary, as it can be very well seen on  Fig.8. \\

\begin{center}
               \includegraphics[width=4cm]{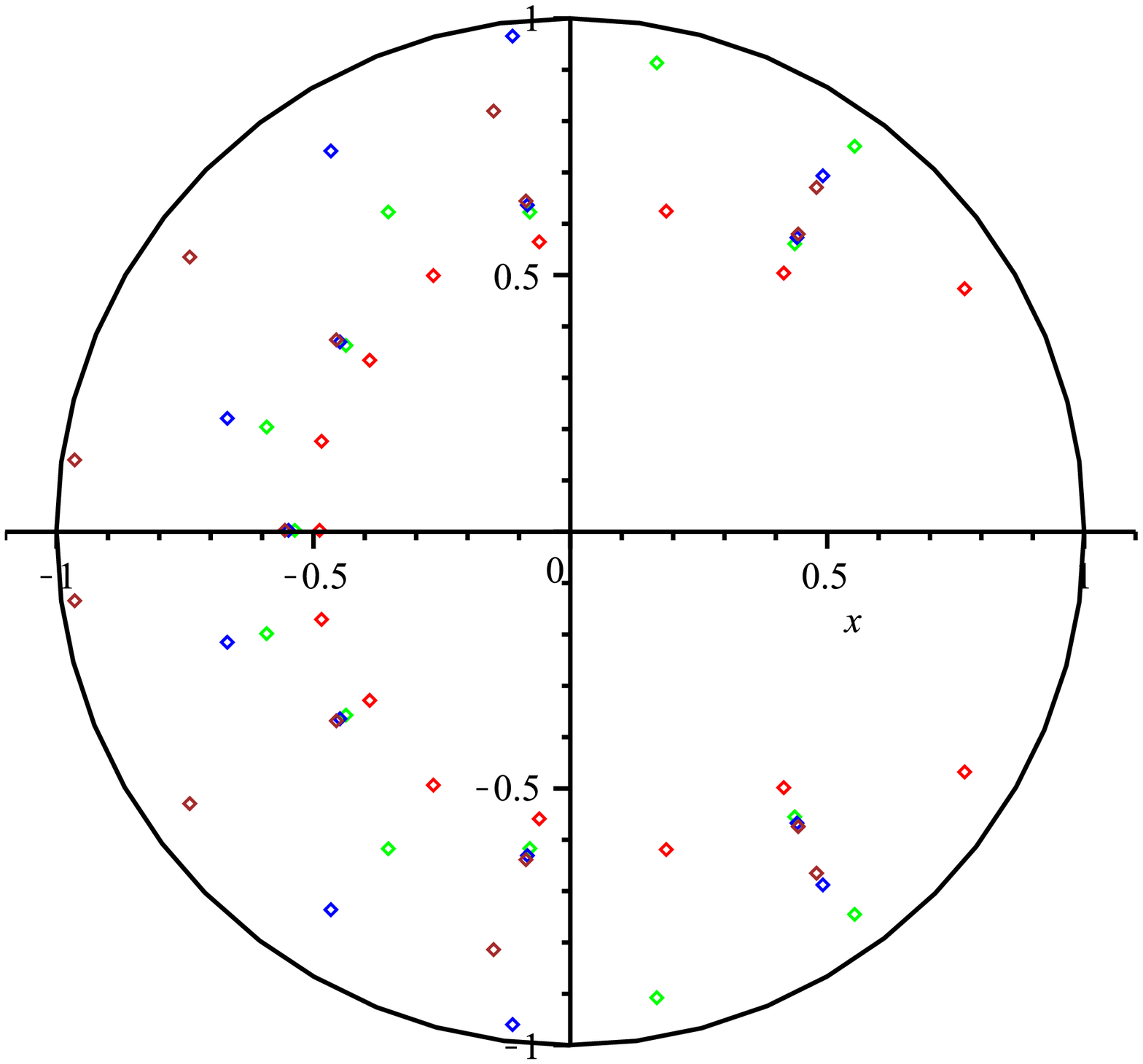}\hspace{2cm}
               \includegraphics[width=4cm]{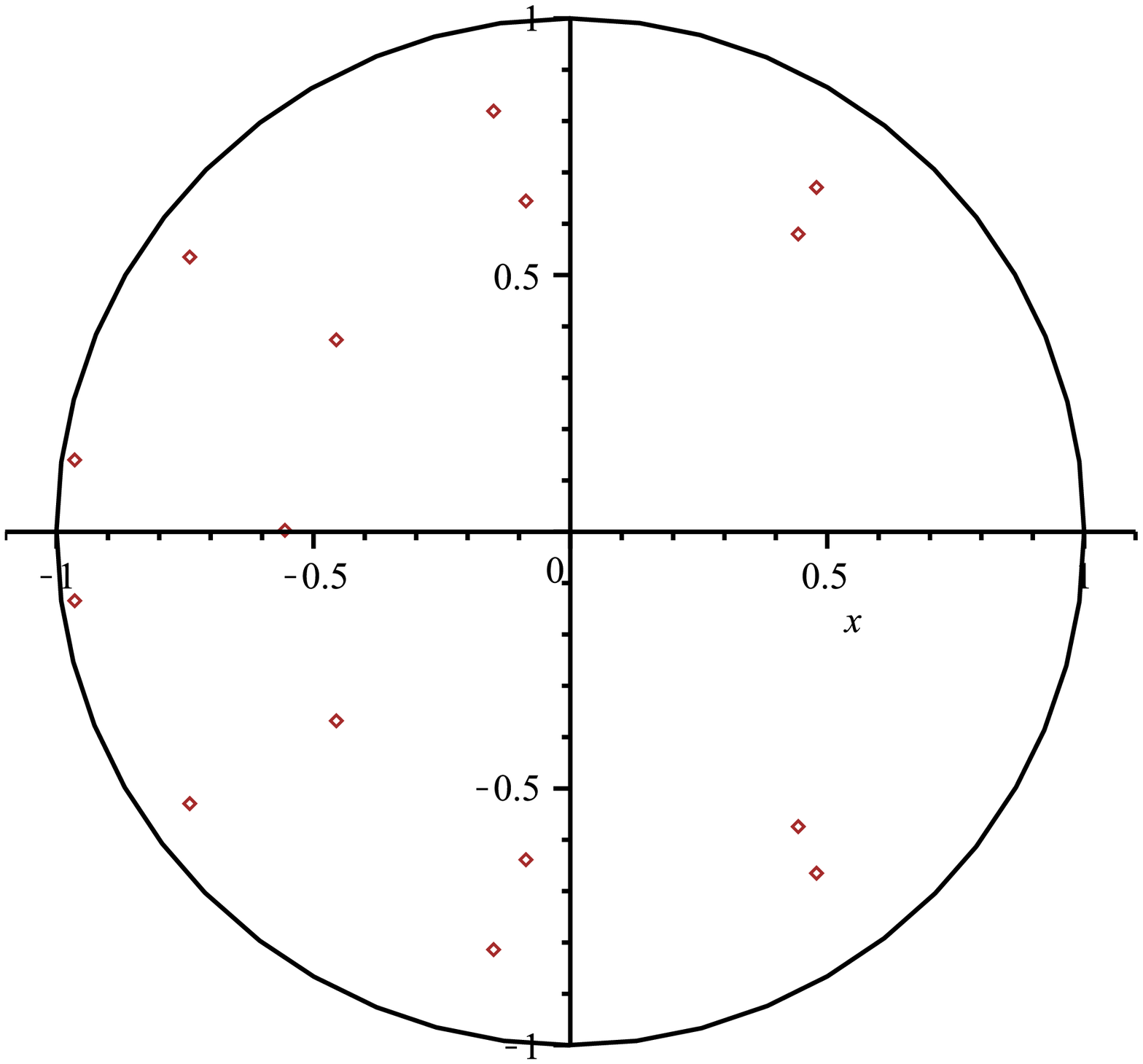}

   \hspace{-1cm}             Fig.7 T=2, $\mu=-2,-20,-40, -63$. Fig.8 T=2, $\mu=-63$
\end{center}

In the case of real multipliers and $T=3, N=8$ the critical value $|I_N^{(3)}|^{-1}=89.72584369...$ Thus, the roots  corresponding the choice $\mu=-2$ (red), $\mu=-30$ (green), $\mu=-60$ (blue) are separated from the unit circle on the Fig.9 while $\mu=-89$ (brown) is close to the critical value and the corresponding roots are close to the boundary, as it can be very well seen on the Fig.10. \\

\begin{center}
               \includegraphics[width=4cm]{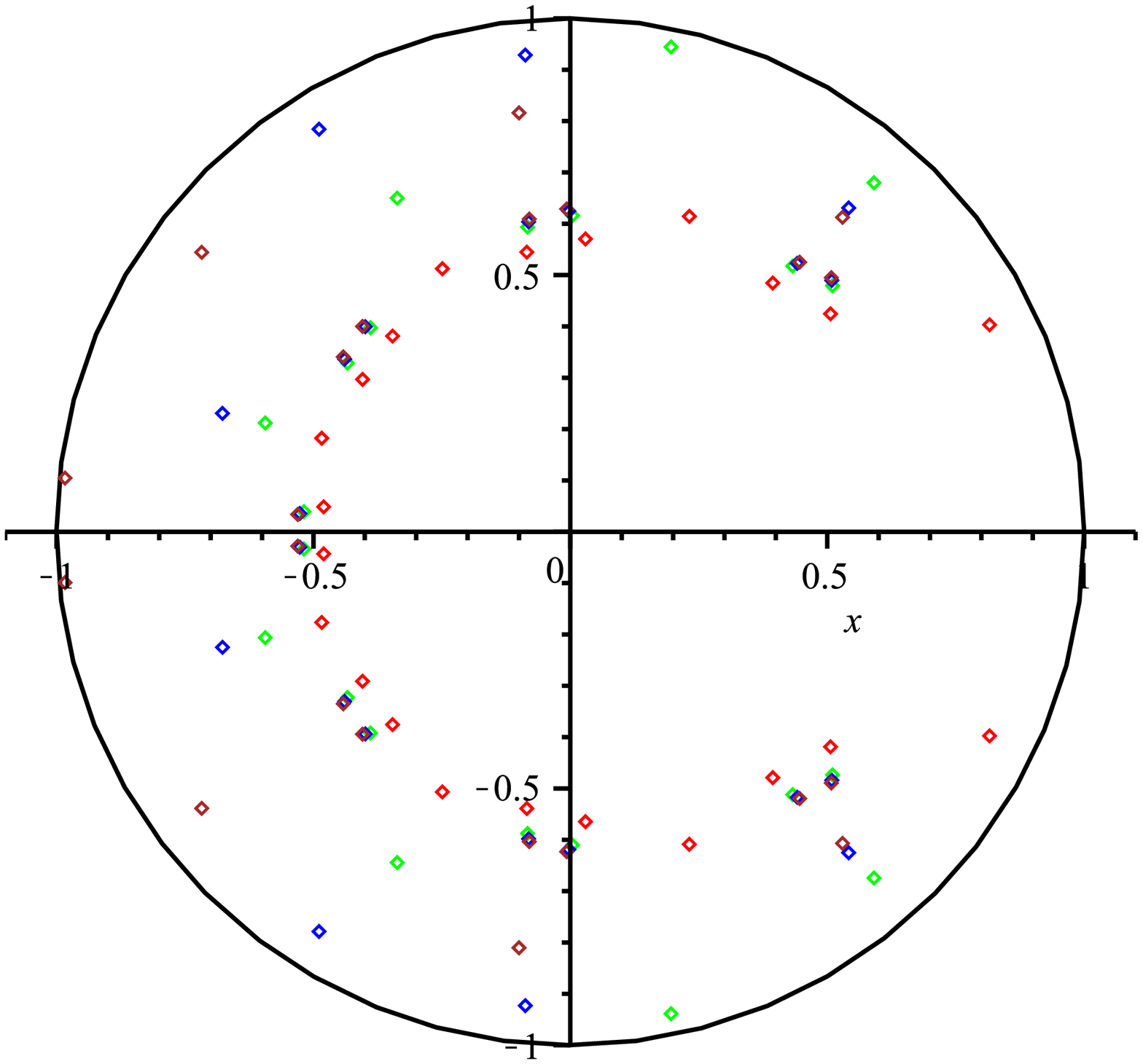}\hspace{2cm}
               \includegraphics[width=4cm]{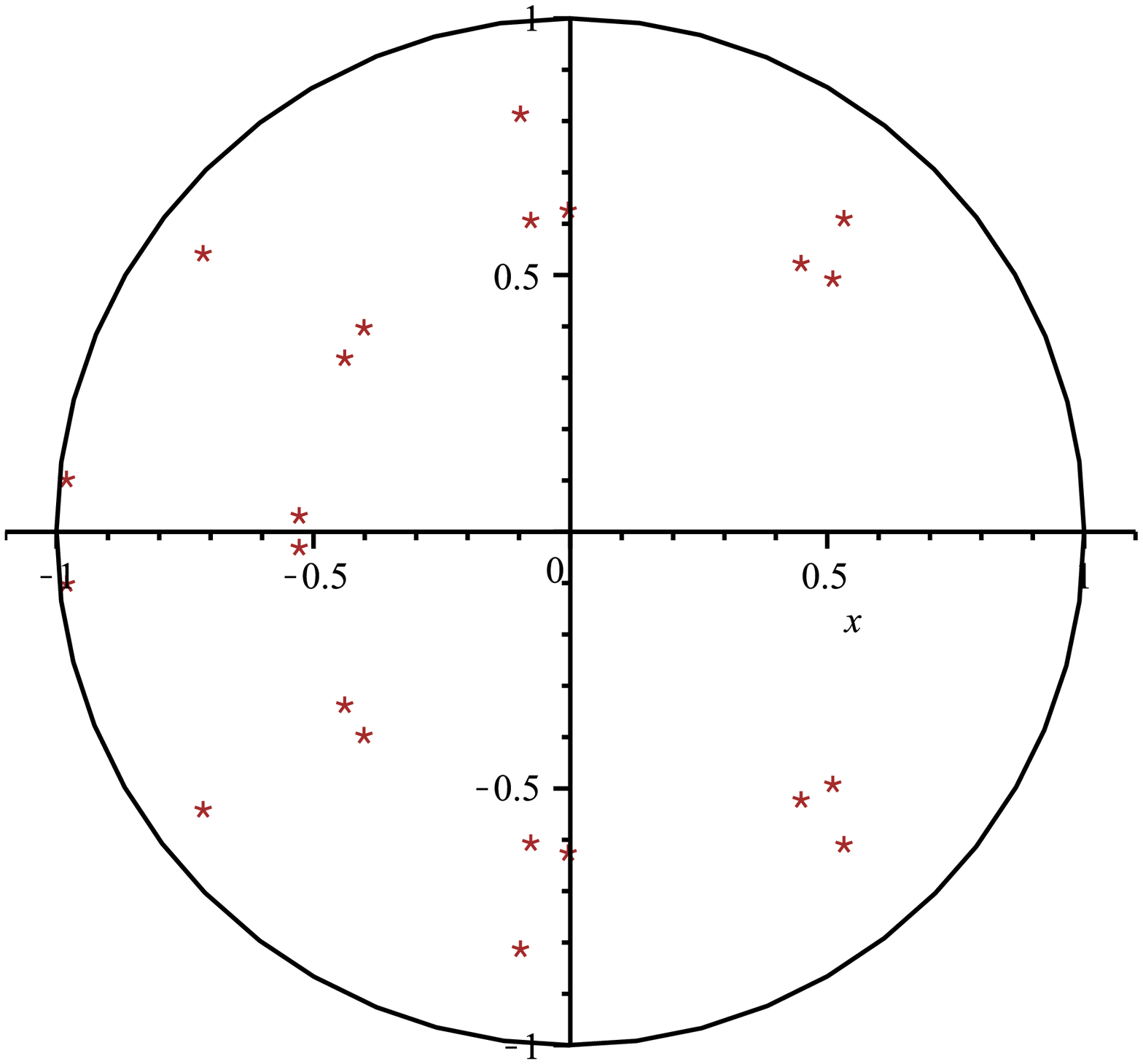}

   \hspace{-1cm}             Fig.9 T=3, $\mu=-2,-30,-60,-89$. Fig.10 T=3, $\mu=-89$
\end{center}

      In the case of complex multipliers and $T=1, N=8$ the critical value $2\cdot R=8$ or $R=4$ (see \eqref{optmuc}). Thus, the roots  corresponding to the choice $\mu=-4+4\exp(0.3\pi i)$ (red),
$\mu=-4+4\exp(0.5\pi i)$ (green), $\mu=-4+4\exp(0.7\pi i)$ (blue), $\mu=-4+4\exp(0.9\pi i)$ (brown) all correspond to the critical value of $R$ and there are roots of each color close to the boundary on the Fig.11. Especially it is very well visible on  Fig.12 where the case $\mu=-4+4\exp(0.9\pi i)$ is illustrated.\\

\begin{center}
               \includegraphics[width=4cm]{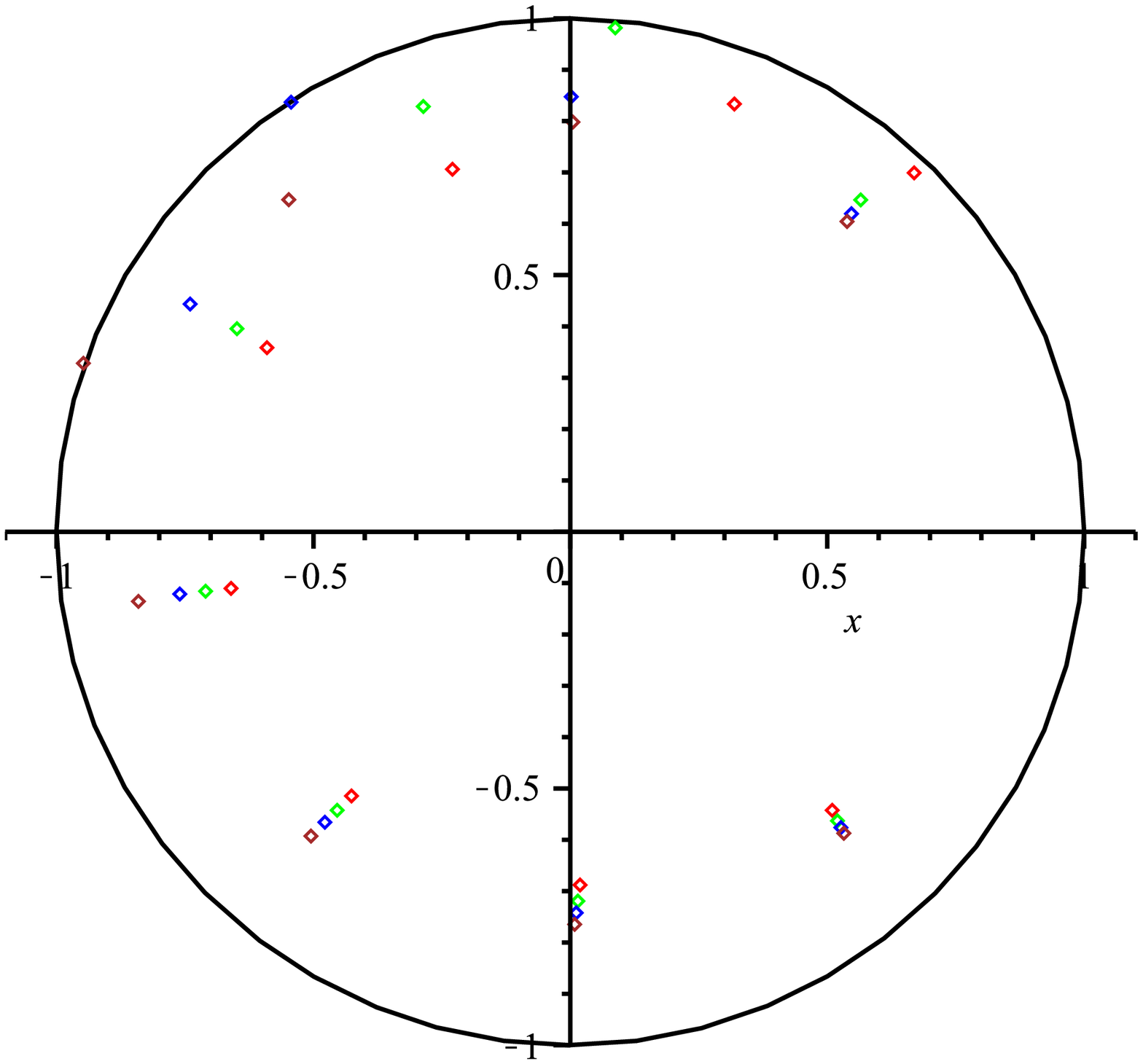}\hspace{2cm}
               \includegraphics[width=4cm]{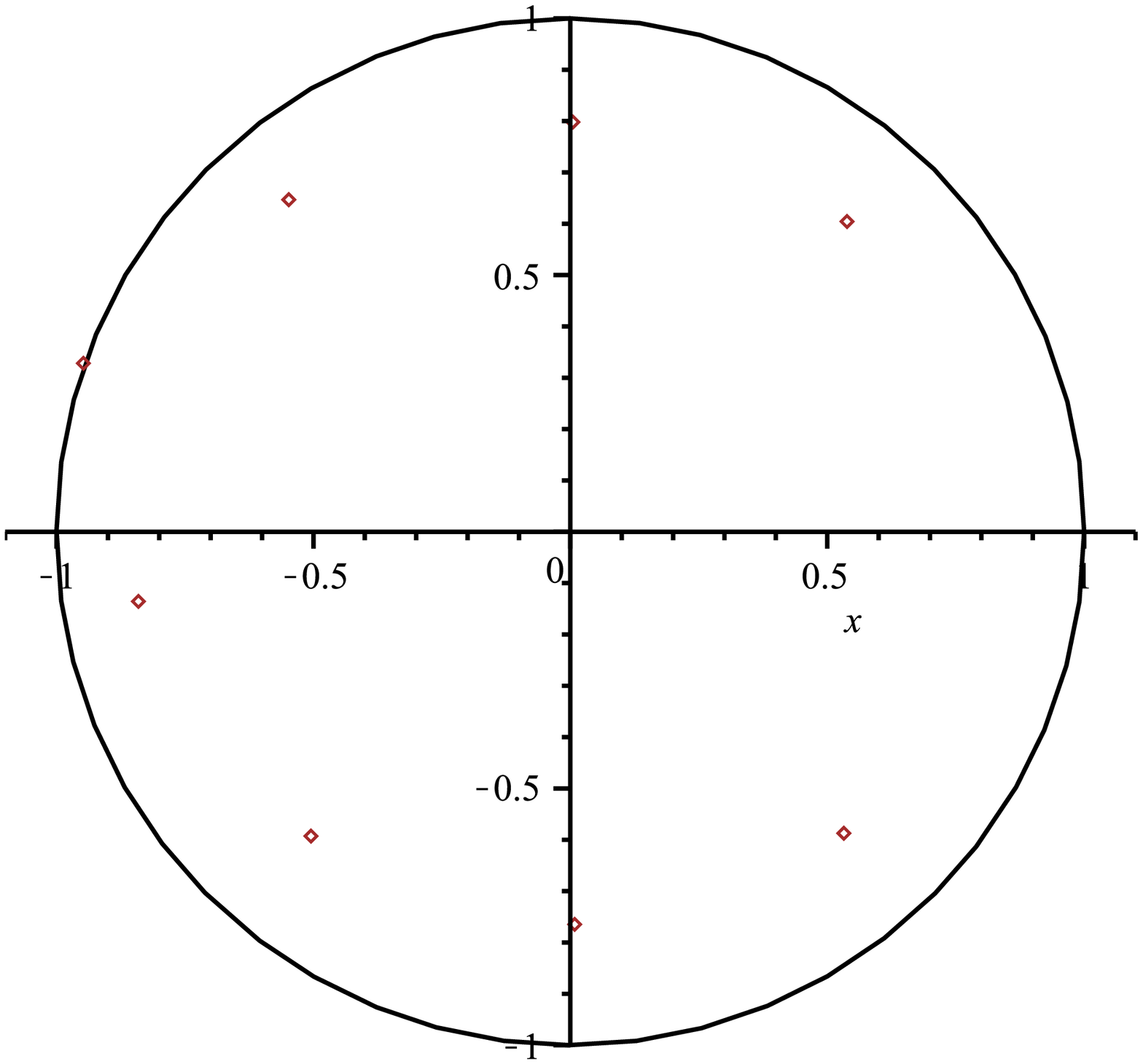}

   \hspace{-1cm}             Fig.11 T=1, $\mu$ complex.  Fig.12 T=1, $\mu=-4+4\exp(0.9\pi i)$
\end{center}

      In the case of complex multipliers and $T=2, N=8$ the critical value $2\cdot R=10.48682166...$  Thus, the roots  corresponding to the choice $\mu=-5.25+5.25\exp(0.3\pi i)$ (red),
$\mu=-5.25+5.25\exp(0.5\pi i)$ (green), $\mu=-5.25+5.25\exp(0.7\pi i)$ (blue), $\mu=-5.25+5.25\exp(0.9\pi i)$ (brown) all correspond to the critical value of $R$ and there are roots of each color close to the boundary on the Fig.13. Especially it is very well visible on Fig.14 where the case $\mu=-5.25+5.25\exp(0.9\pi i)$ is illustrated.\\

\begin{center}
               \includegraphics[width=4cm]{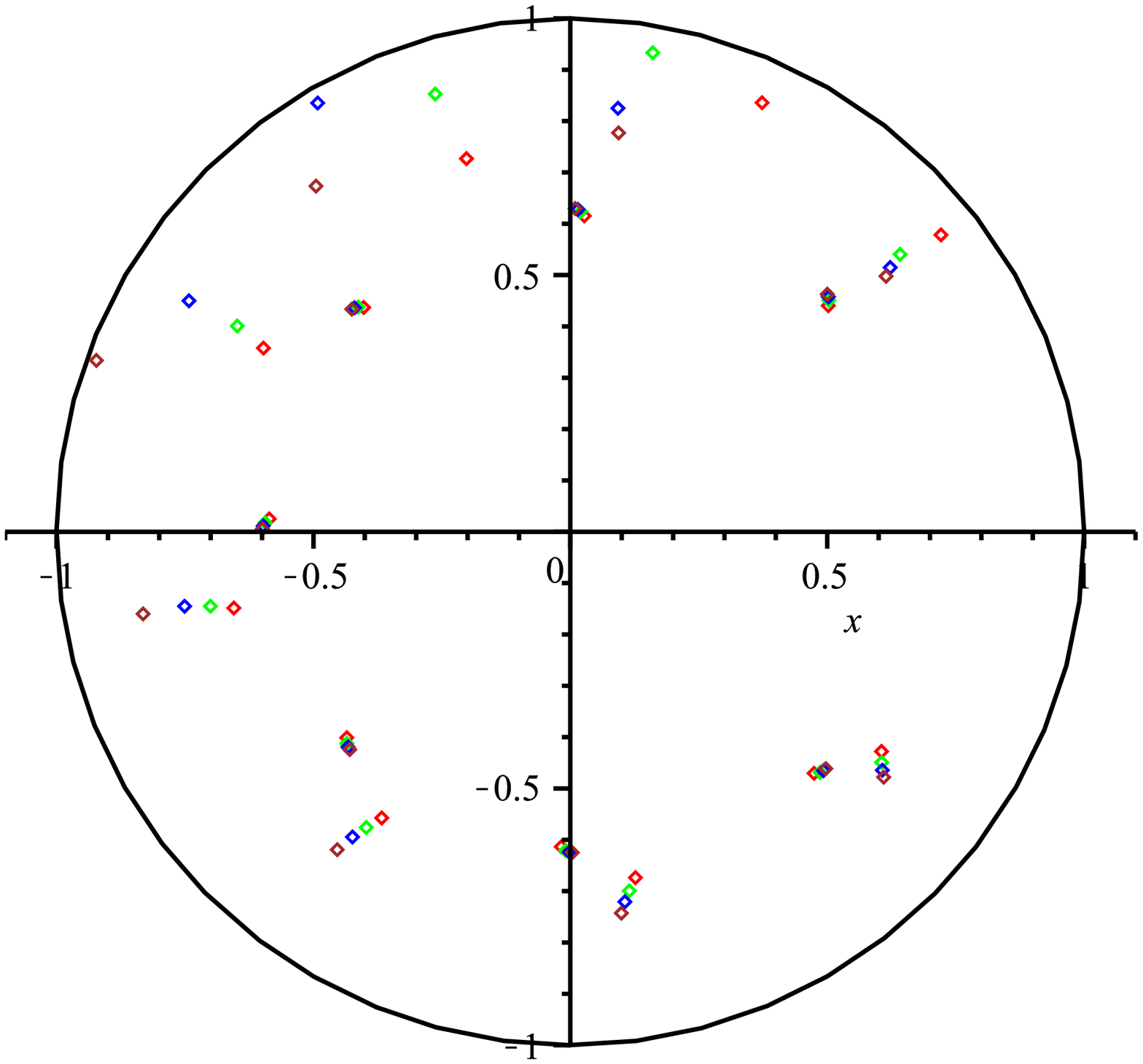}\hspace{2cm}
               \includegraphics[width=4cm]{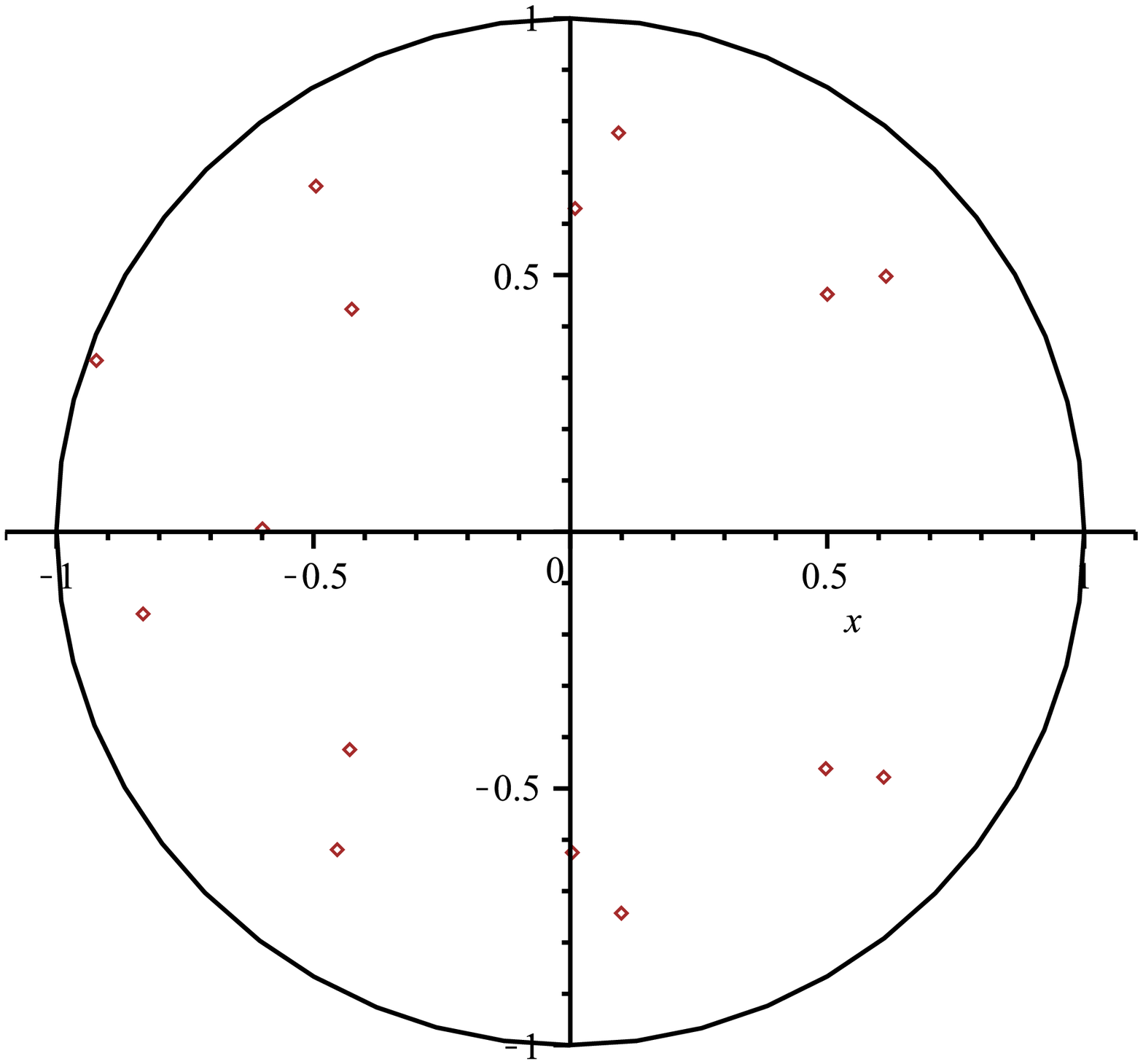}

   \hspace{-1cm}             Fig.13 T=2, $\mu$ complex.  Fig.14 T=2, $\mu=-5.25+5.25\exp(0.7\pi i)$
\end{center}

In the case of complex multipliers and $T=3, N=8$ the critical value $2\cdot R=11.79242673...$  In this case we choose the multiplier even smaller in absolute value than critical by choosing 5.8 instead of 5.9 below. However, even in this case one can observe the existence of roots of each color close to the boundary on Fig.15. There $\mu=-5.9+5.8\exp(0.3\pi i)$ (red), $\mu=-5.9+5.8\exp(0.5\pi i)$ (green), $\mu=-5.9+5.8\exp(0.7\pi i)$ (blue), $\mu=-5.9+5.8\exp(0.9\pi i)$ (brown). It is very well visible on the Fig.16 with $\mu=-5.9+5.8\exp(0.9\pi i)$  that there are roots almost on the boundary.\\

\begin{center}
               \includegraphics[width=4cm]{RootT3.eps}\hspace{2cm}
               \includegraphics[width=4cm]{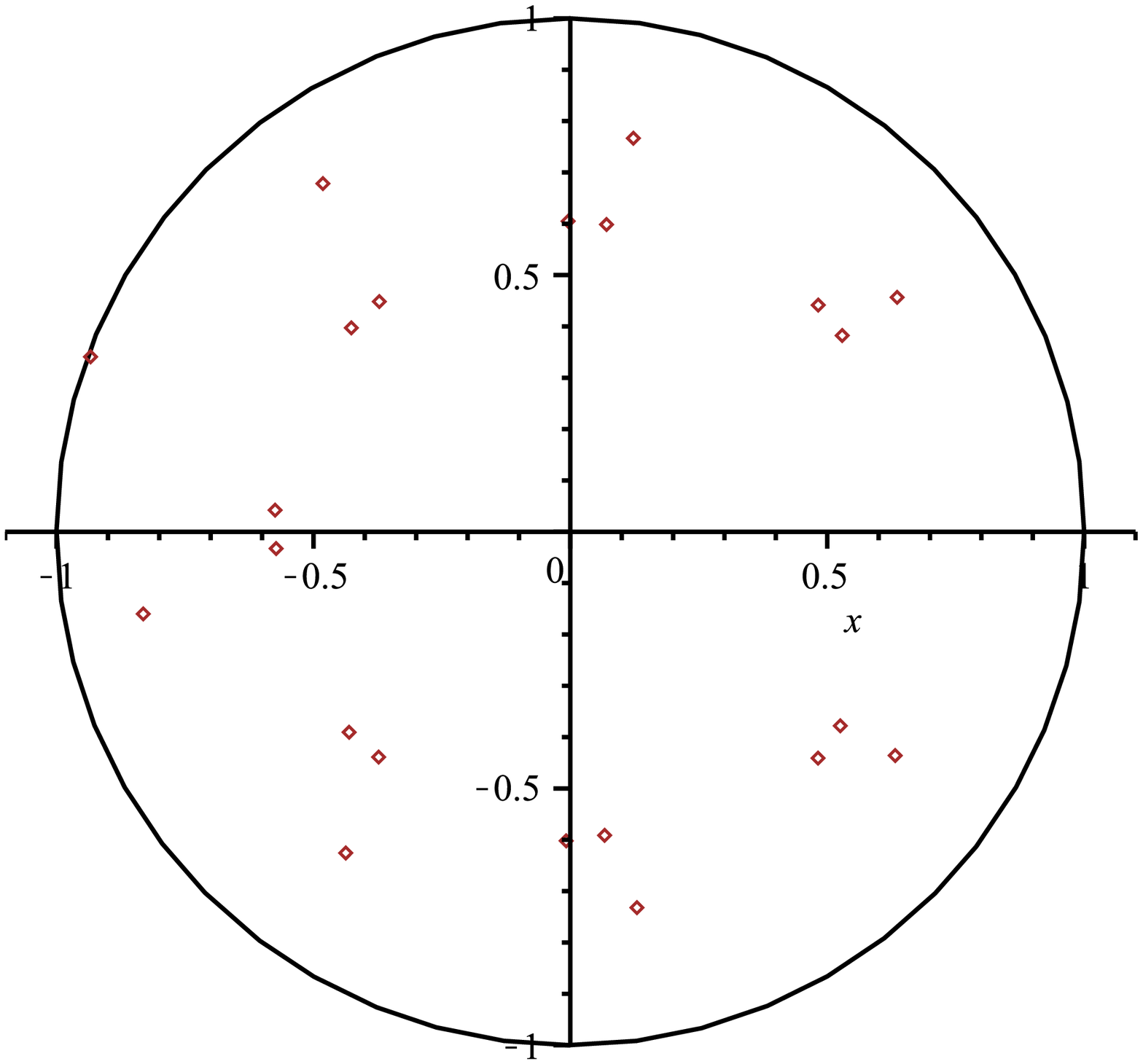}

   \hspace{-1cm}             Fig.15 T=3,  $\mu$ complex.  Fig.16 T=3, $\mu=-5.9+5.8\exp(0.7\pi i)$
\end{center}

\subsection{Root location. Modified polynomials.}

\begin{center}
               \includegraphics[width=4cm]{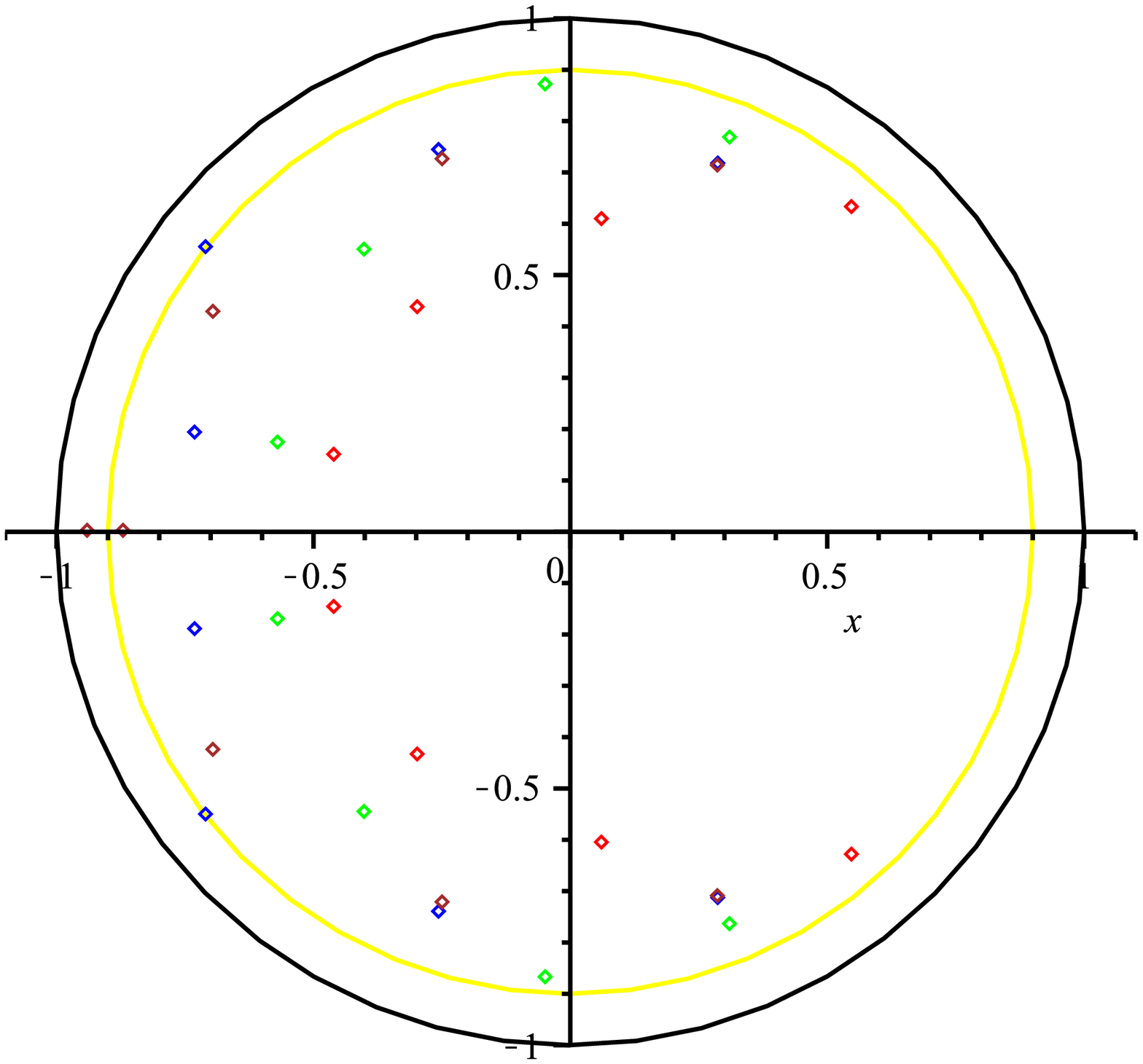}\hspace{2cm}
               \includegraphics[width=4cm]{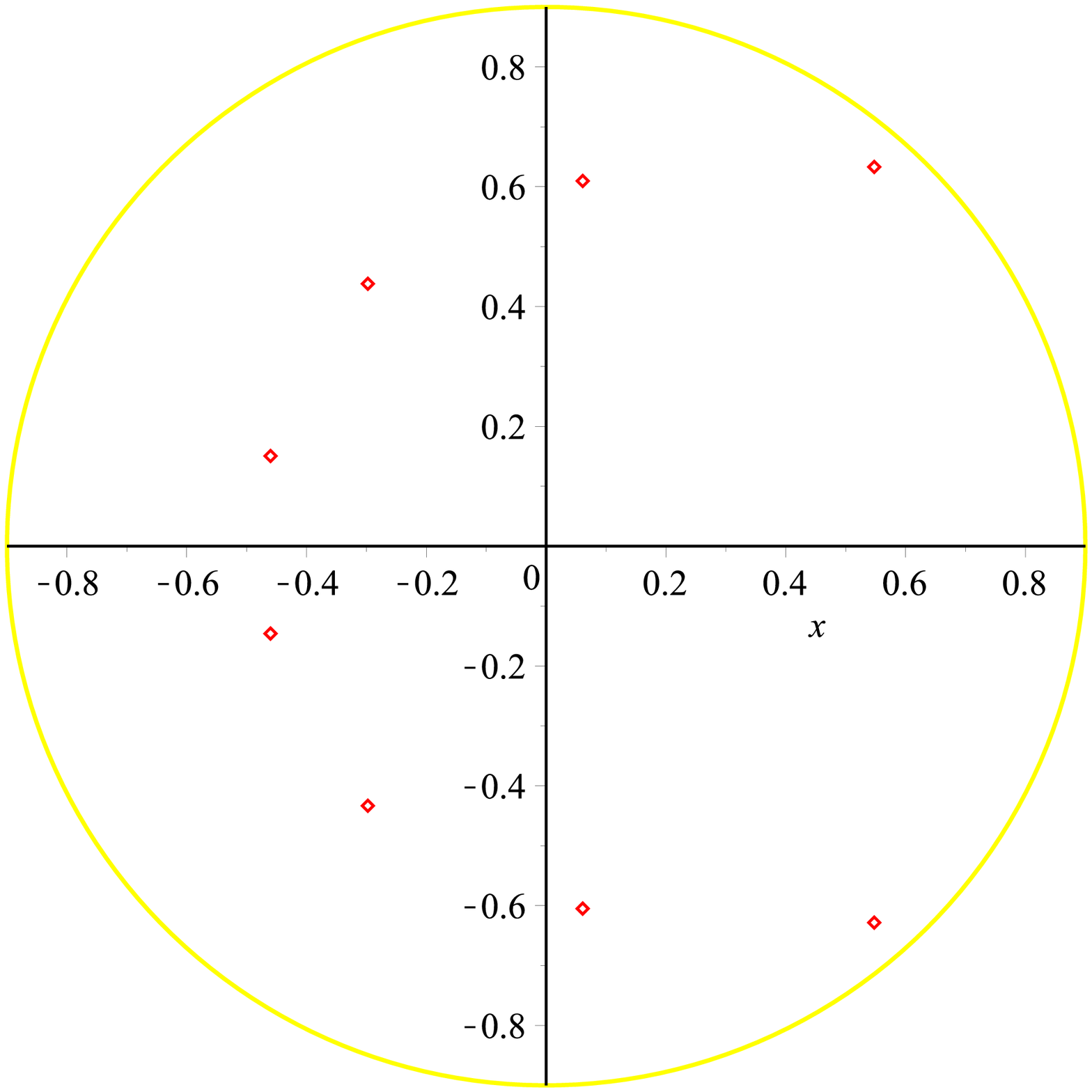}

   \hspace{-1cm}             Fig.17 T=1,  $\mu= -2, -10, -20, -22.2.$  Fig.18 T=1, $\rho=0.9, \mu=-2$
\end{center}

In this case $T=1, N=8, \rho=0.9$ we have the estimate
$$
\mu^*<(a_1^{(8)}0.9+...+a_8^{(8)}0.9^8)\cot^2\frac\pi{18}
=22.17436354...
$$
On  Fig.17  $\mu= -2$ (red),  $\mu=-10$ (green), $\mu=-20$ (blue) and $\mu=-22.2$ (brown). As one can observe all the roots except  brown lie inside a disc of radius 0.9. Especially it is very well visible on  Fig.18 where the case $\mu=-2$ is displayed.
If multiplier is large then even the modified polynomial can have roots outside a disc of radius $0.9.$ It happen to the brown roots because $\mu=-22.2$ is slightly smaller then the critical value for parameters $\rho=0.9, N=8$ which is $-22.17.$
Note, that in this case $\cot^2\frac\pi{18}=32.16343748...$ therefore the brown root is inside the unit circle.

  %In this case the value $\hat\mu^*=$

Let us consider the real case $T=2, N=8, \rho=0.9$ In this situation the critical value for the multipliers is -37.71670343...
As one can observe from the Fig.19 all the roots except  brown
($\mu=-37.8$) are inside a disc of radius 0.9. Especially it is very well visible on  Fig.20 where the case $\mu=-15$ is displayed. One of the brown roots lies outside the circle of radius 0.9 because the brown multiplier  $\mu=-37.8$ is slightly smaller then the critical value -37.71.\\

\begin{center}
               \includegraphics[width=4cm]{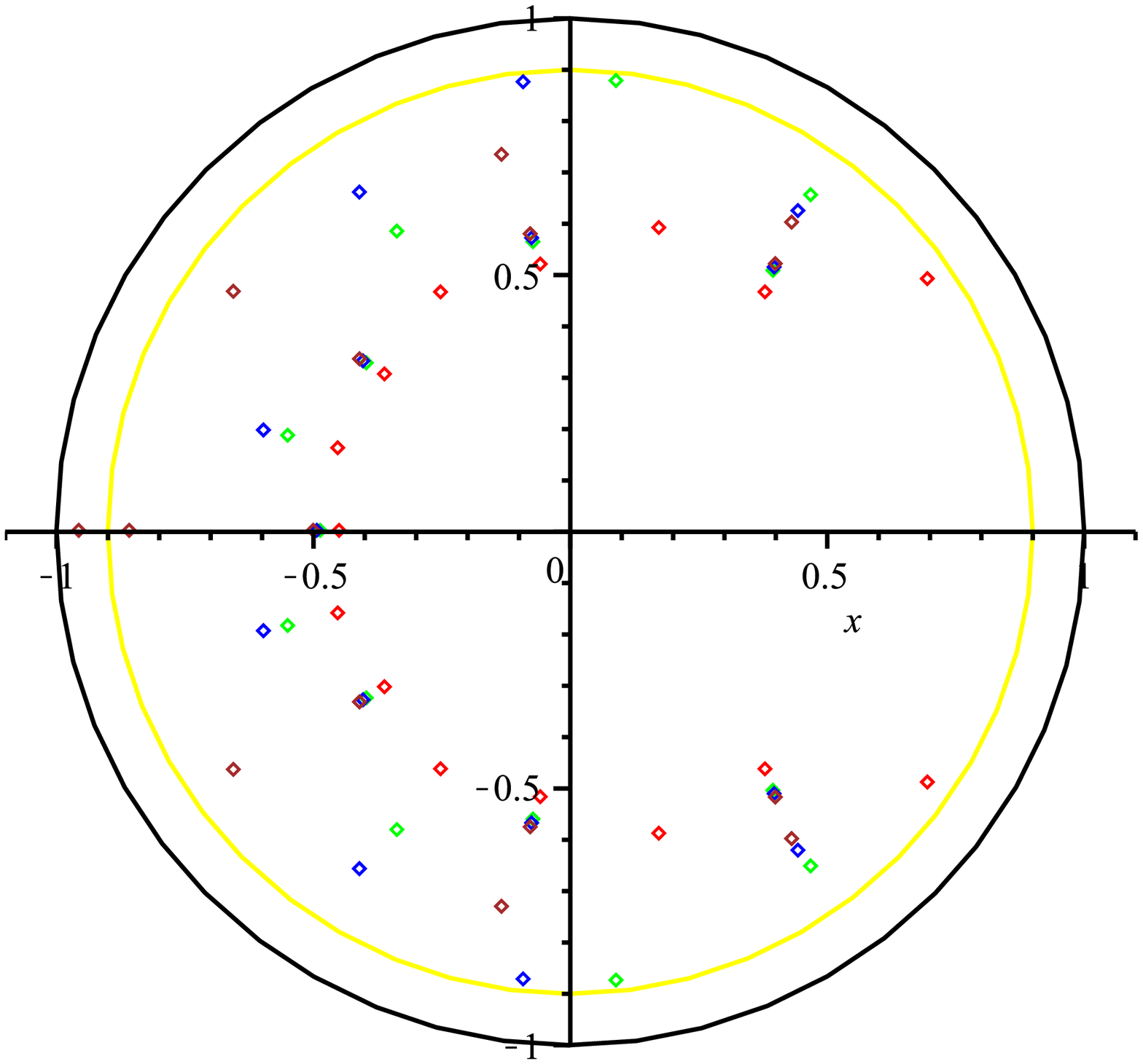}\hspace{2cm}
               \includegraphics[width=4cm]{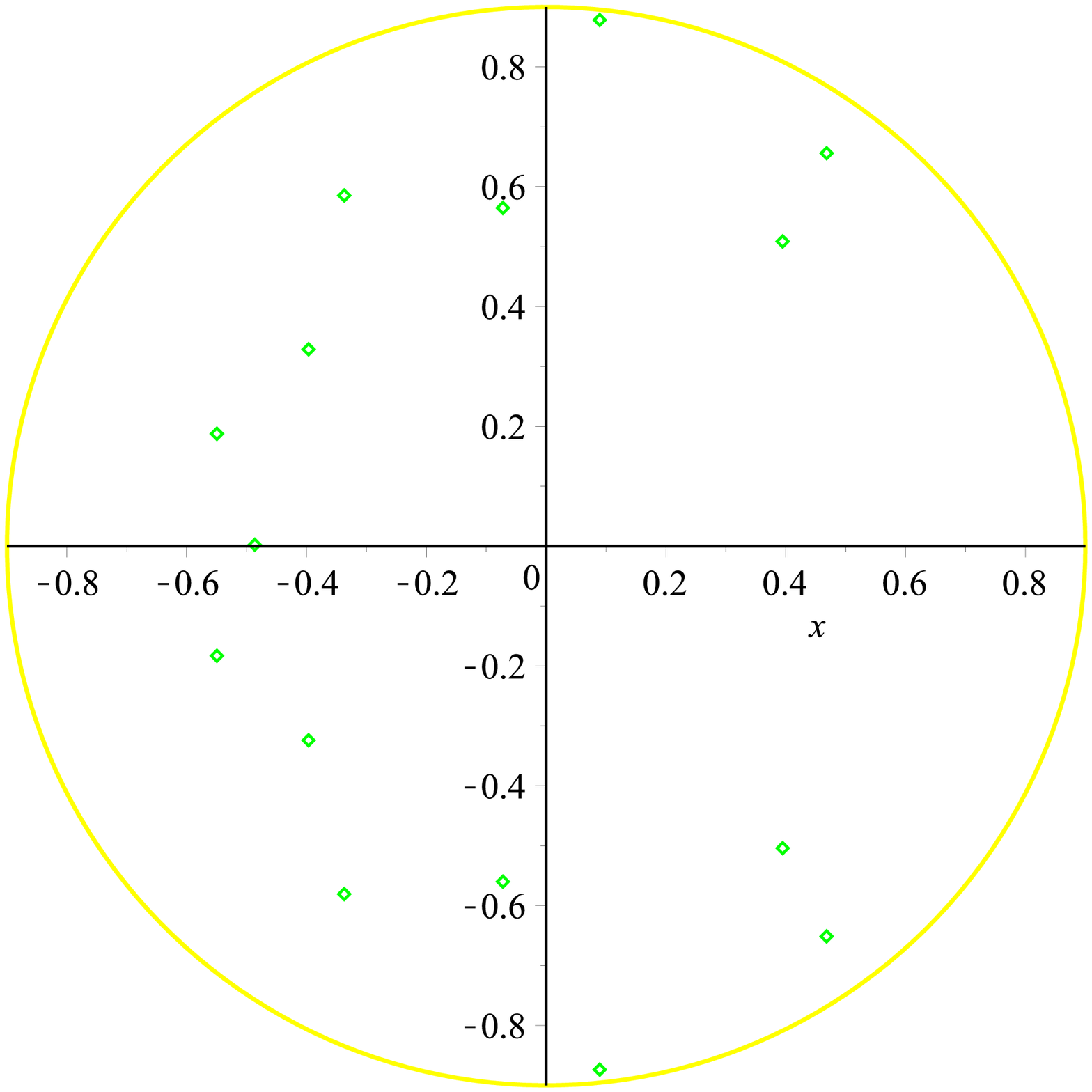}

   \hspace{-1cm}             Fig.19 T=2,  $\mu= -2, -15, -23, -37.8.$  Fig.20 T=2, $\rho=0.9, \mu=-15$
\end{center}

Now, let us consider the real case $T=3, N=8, \rho=0.9$ In this situation the critical value for the multipliers is -47.82046491...
As one can observe from Fig.21 all the roots except  brown
($\mu=-48$) are inside a disc of radius 0.9. Especially it is very well visible on the Fig.22 where the case $\mu=-2$ is displayed. One of the brown roots lies outside the circle of radius 0.9 because the brown multiplier  $\mu=-48.8$ is slightly smaller then the critical value -47.8.\\

\begin{center}
               \includegraphics[width=4cm]{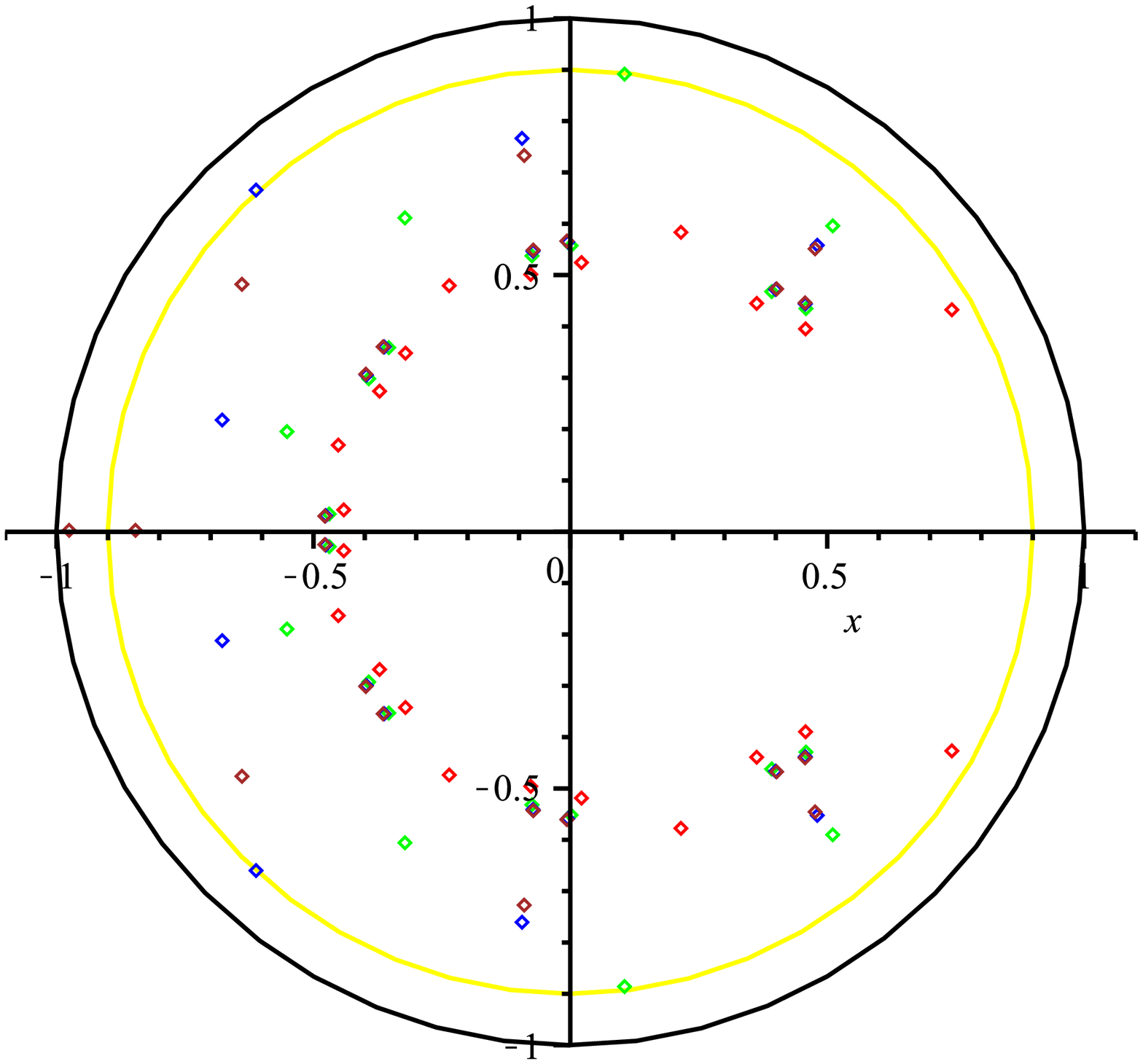}\hspace{2cm}
               \includegraphics[width=4cm]{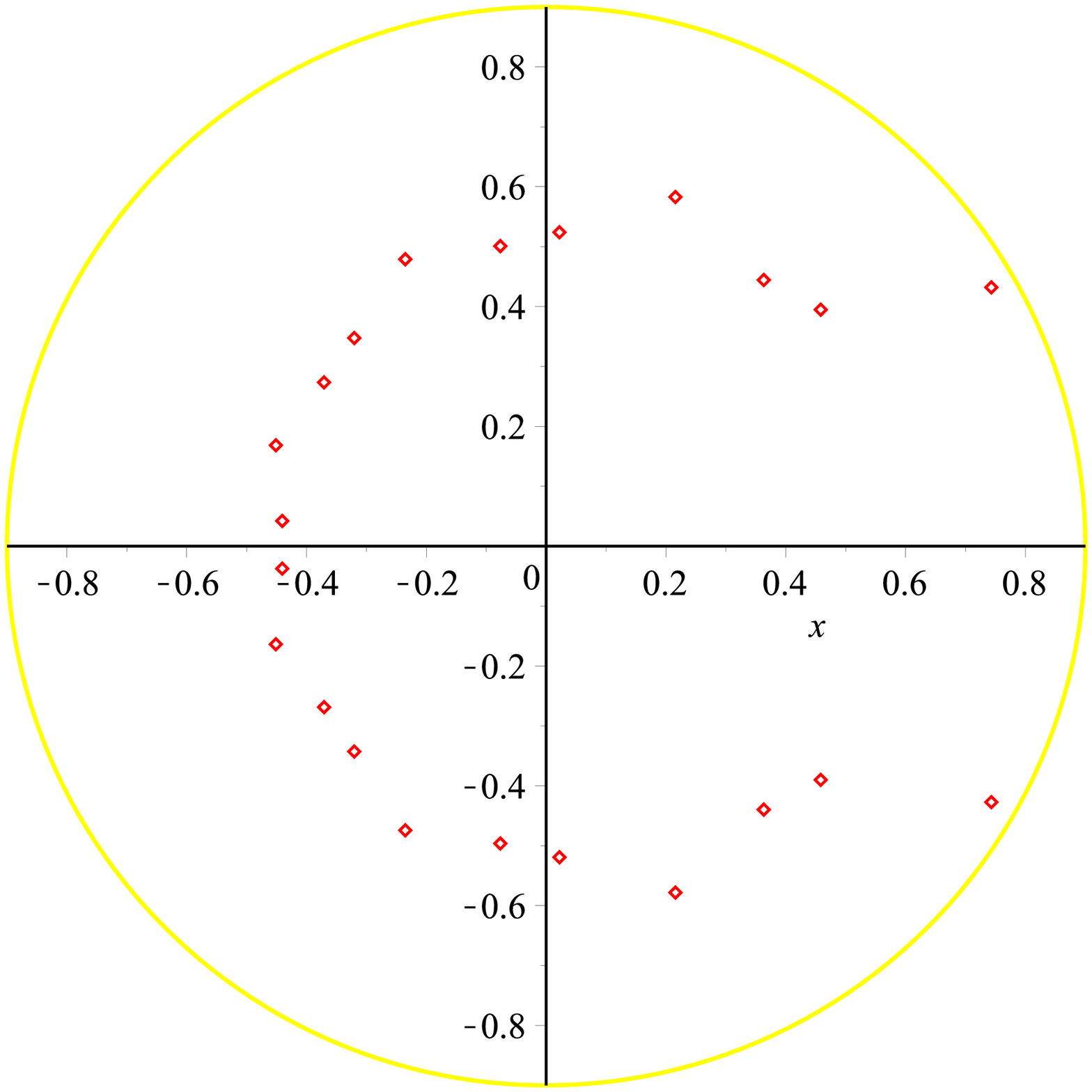}

   \hspace{-1cm}             Fig.21 T=3,  $\mu= -2, -20, -40, -48.$  Fig.22 T=3, $\rho=0.9, \mu=-2$
\end{center}

Now, let us consider the complex case $T=1, N=8, \rho=0.9$ In this situation the critical value for the multipliers is -5.748409779...
As one can observe from  Fig.23 all the roots $\mu=-2.9+2.9\exp(\frac{\pi i}5),$ $\mu=-2.9+2.9\exp(\frac{2\pi i}5),$
$\mu=-2.9+2.9\exp(\frac{4\pi i}5)$
except  brown
($\mu=-6.5$) are inside a disc of radius 0.9. Especially it is very well visible on Fig.24 where the case $\mu=-2.9+2.9\exp(\frac{\pi i}5)$ is displayed. One of the brown roots lies outside the circle of radius 0.9 because the brown multiplier  $\mu=-6.5$ is slightly smaller then the critical value -5.7 although still inside the unit disc.\\

\begin{center}
               \includegraphics[width=4cm]{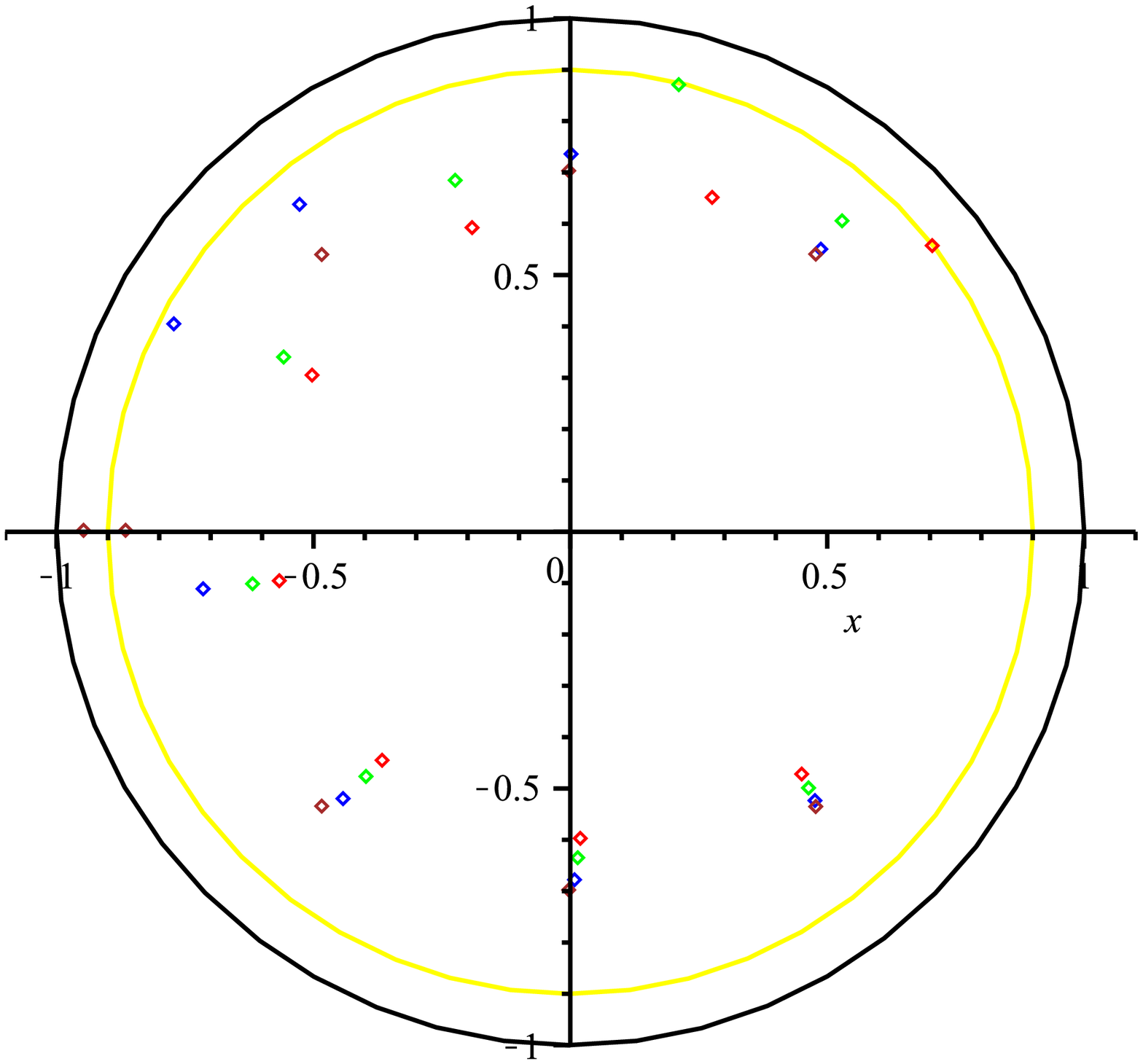}\hspace{2cm}
               \includegraphics[width=4cm]{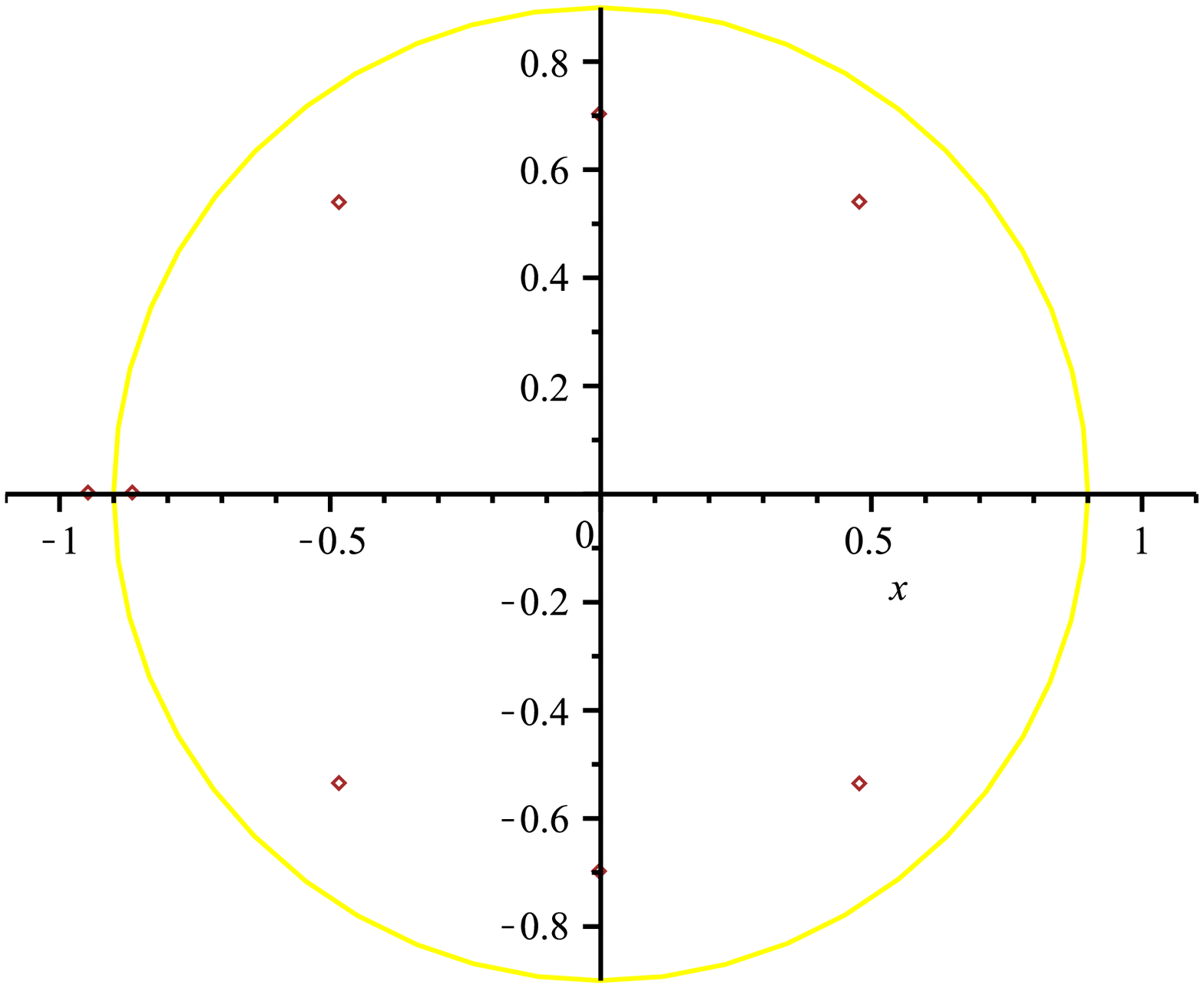}

   \hspace{-1cm}             Fig.23 T=1,  $\mu$ complex.  Fig.24 T=1, $\rho=0.9, \mu=-5.9+5.8\exp(0.7\pi i)$
\end{center}

Further, let us consider the complex case $T=2, N=8, \rho=0.9$ In this situation the critical value for the multipliers is -6.871373952...
As one can observe from  Fig.25 all the roots $\mu=-3.4+3.4\exp(\frac{\pi i}5),$ $\mu=-2.9+2.9\exp(\frac{2\pi i}5),$
$\mu=-2.9+2.9\exp(\frac{2.8\pi i}5)$
except  brown
($\mu=-3.6+3.6\exp(\frac{3\pi i}5)$) are inside a disc of radius 0.9. Especially it is very well visible on Fig.26 where the case $\mu=-3.6+3.6\exp(\frac{3\pi i}5)$ is displayed. One of the brown roots lies outside the circle of radius 0.9 because the brown multiplier  $\mu=-3.6+3.6\exp(\frac{3\pi i}5)$ in absolute value is slightly larger then the critical value 6.87, however still inside the unit disc.\\

\begin{center}
               \includegraphics[width=4cm]{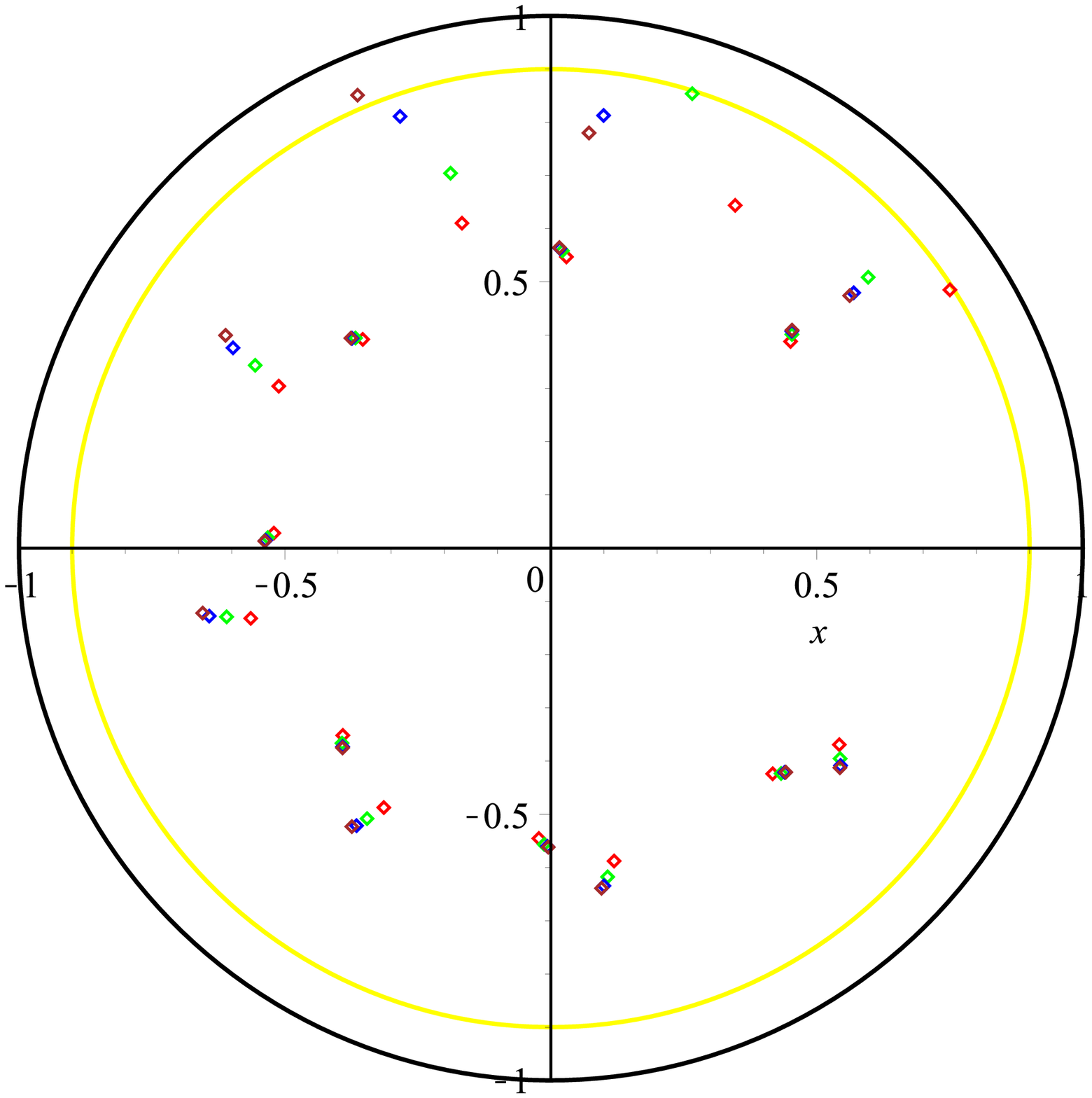}\hspace{2cm}
               \includegraphics[width=4cm]{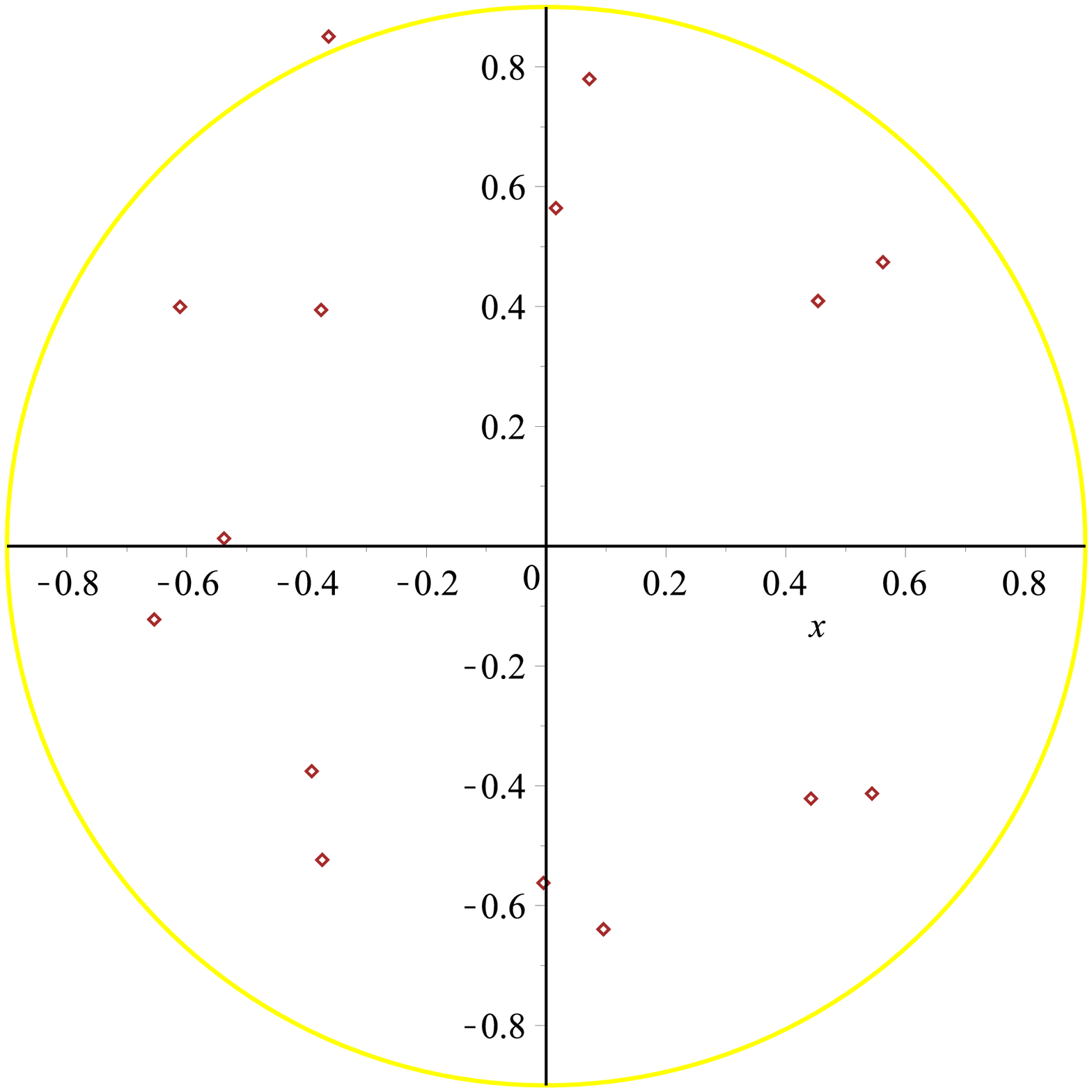}

   \hspace{-1cm}             Fig.25 T=2,  $\mu$ complex.  Fig.26 T=2, $\mu=-3.6+3.6\exp(\frac{3\pi i}5)$
\end{center}

Finally, let us consider the complex case $T=3, N=8, \rho=0.9$ In this situation the critical value for the multipliers is 7.362286563... As one can observe from  Fig.27 all the roots $\mu= -3.7+3.7\exp(\frac{\pi i}5),$ $\mu= -3.7+3.7\exp(\frac{2\pi i}5),$
$\mu= -3.7+3.7\exp(\frac{2.8\pi i}5)$
except  brown
($\mu=-3.9+3.9\exp(\frac{3\pi i}5)$) are inside a disc of radius 0.9. Especially it is very well visible on  Fig.28 where the case $\mu=-3.6+3.6\exp(\frac{3\pi i}5)$ is displayed. One of the brown roots lies outside the circle of radius 0.9 because the brown multiplier  $\mu=-3.9+3.9\exp(\frac{3\pi i}5)$ in absolute value is slightly larger then the critical value 7.36, however still inside the unit disc.\\

\begin{center}
               \includegraphics[width=4cm]{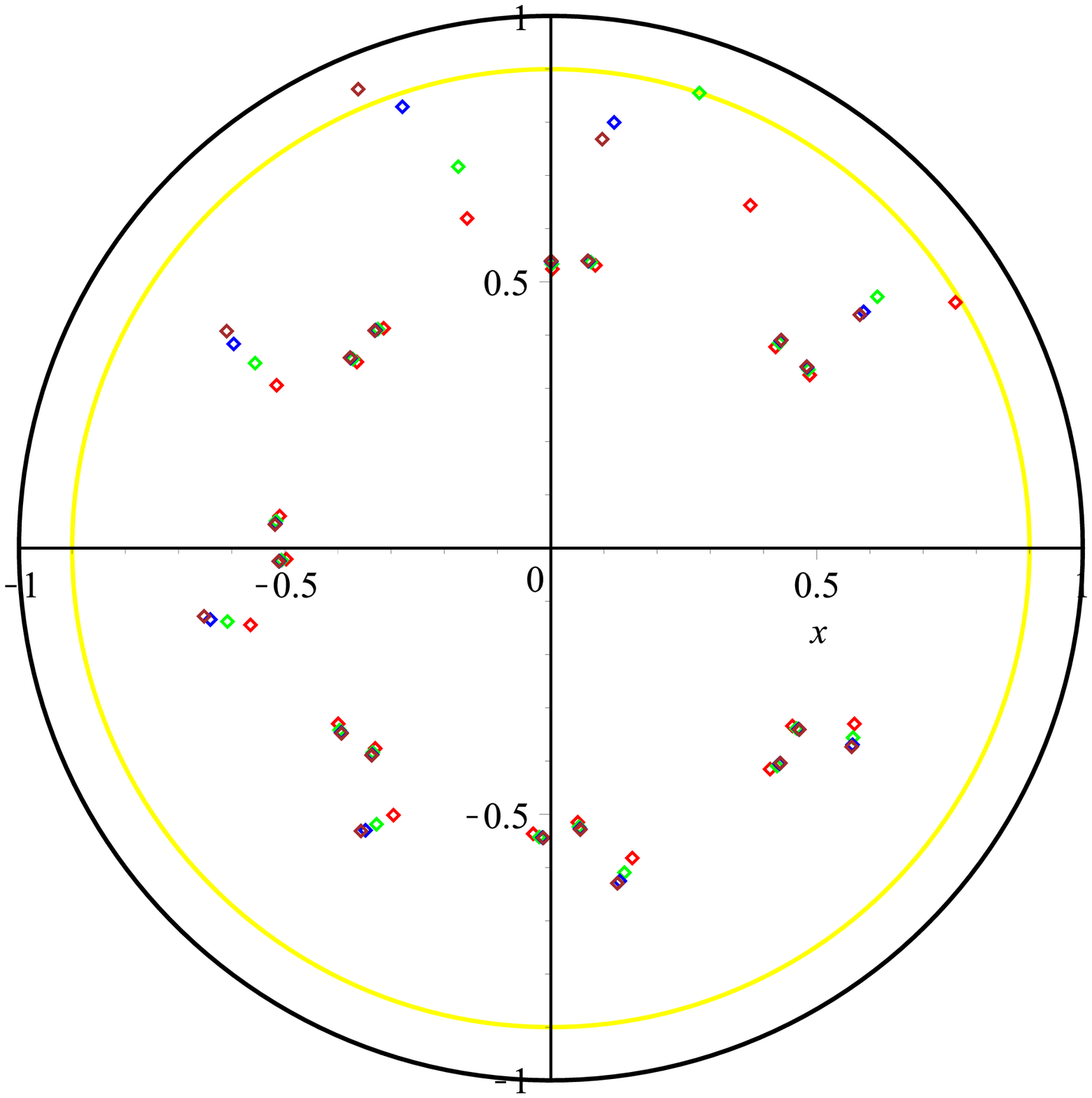}\hspace{2cm}
               \includegraphics[width=4cm]{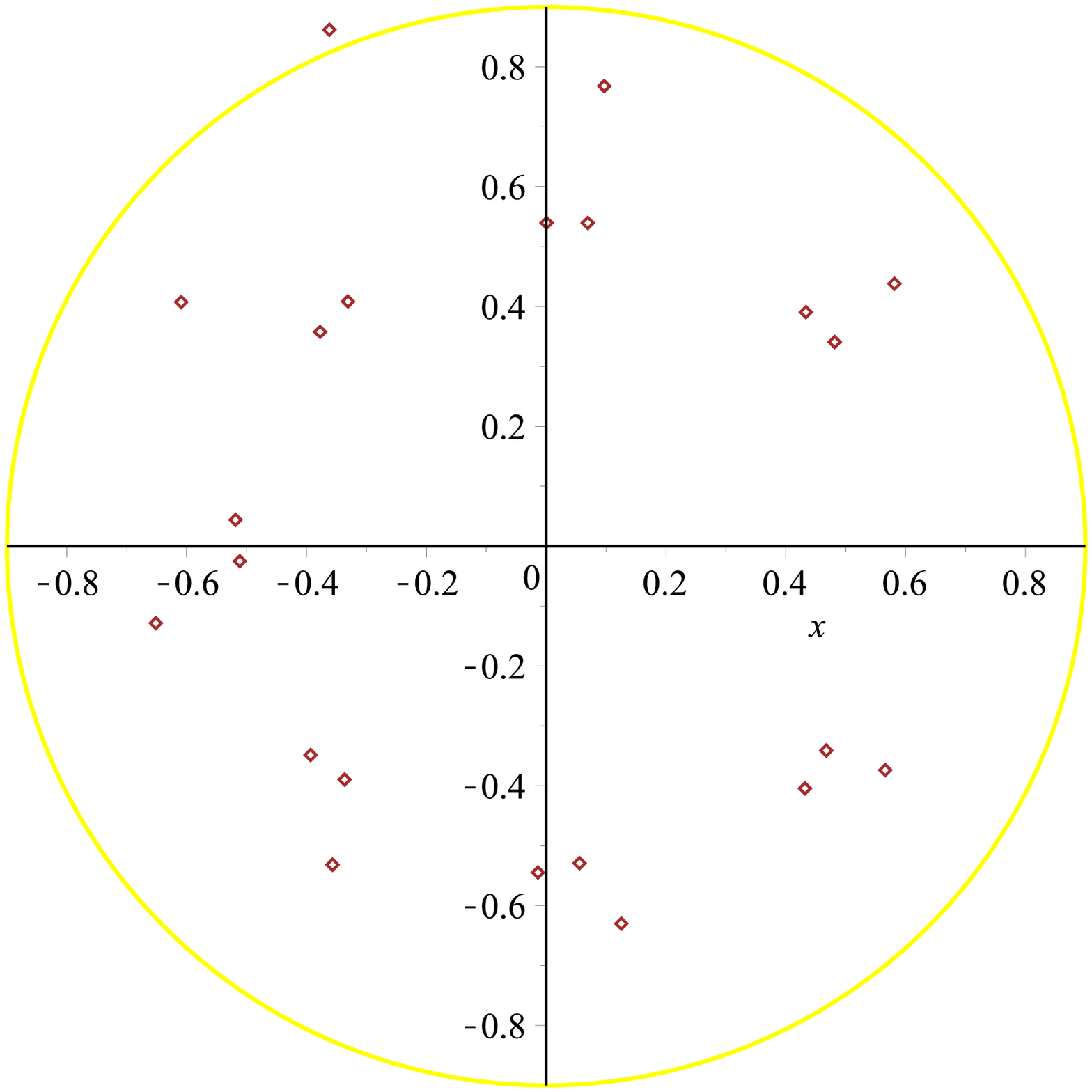}

   \hspace{-1cm}             Fig.27 T=3,  $\mu$ complex.  Fig.28 T=3, $\mu=-3.9+3.9\exp(\frac{3\pi i}5)$.
\end{center}

\section{Examples}

\subsection{Example 1} See \cite{F}. Let $\hat \mu^*<3.$ Let us investigate the problems of constructing the optimal control in the system \eqref{2} for $N=2.$\\

In this case $\hat p(\lambda)=\frac13\lambda+\frac23,$
$\lambda\hat q(\lambda)=\frac23\lambda+\frac13\lambda^2.$ Therefore, $\rho$ should satisfy the inequality $\mu^*<\left(\frac23\rho+\frac13\rho^2 \right)3=2\rho+\rho^2$,
or $\rho>-1+\sqrt{1+\mu^*}.$ For example, for $\hat \mu^*=2$ the minimally possible value $\rho$ is equal to $-1+\sqrt 3\approx 0.73.$ The optimal polynomial is
$$
p_{O}(\lambda)=\frac1{\frac23\frac1\rho+\frac13}
\left(\frac23\frac\lambda\rho+\frac13\right)=\lambda\frac2{2+\rho}+
\frac\rho{2+\rho}.
$$
If we want to increase the rate of convergence to exceed 0.73 we need to chose $N>2$ in the control system \eqref{2}.\\

\subsection{Example 2} The well-known in biology Ally effect is modeling by a bell shaped equation, e.g.
\begin{equation}\label{ally}
x_{n+1}=F(x_n),\quad F(x)=\frac{e^{-5(2x-1)^2}-e^{-5}}{1-e^{-5}}
\end{equation}
\centerline{
\includegraphics[scale=0.25]{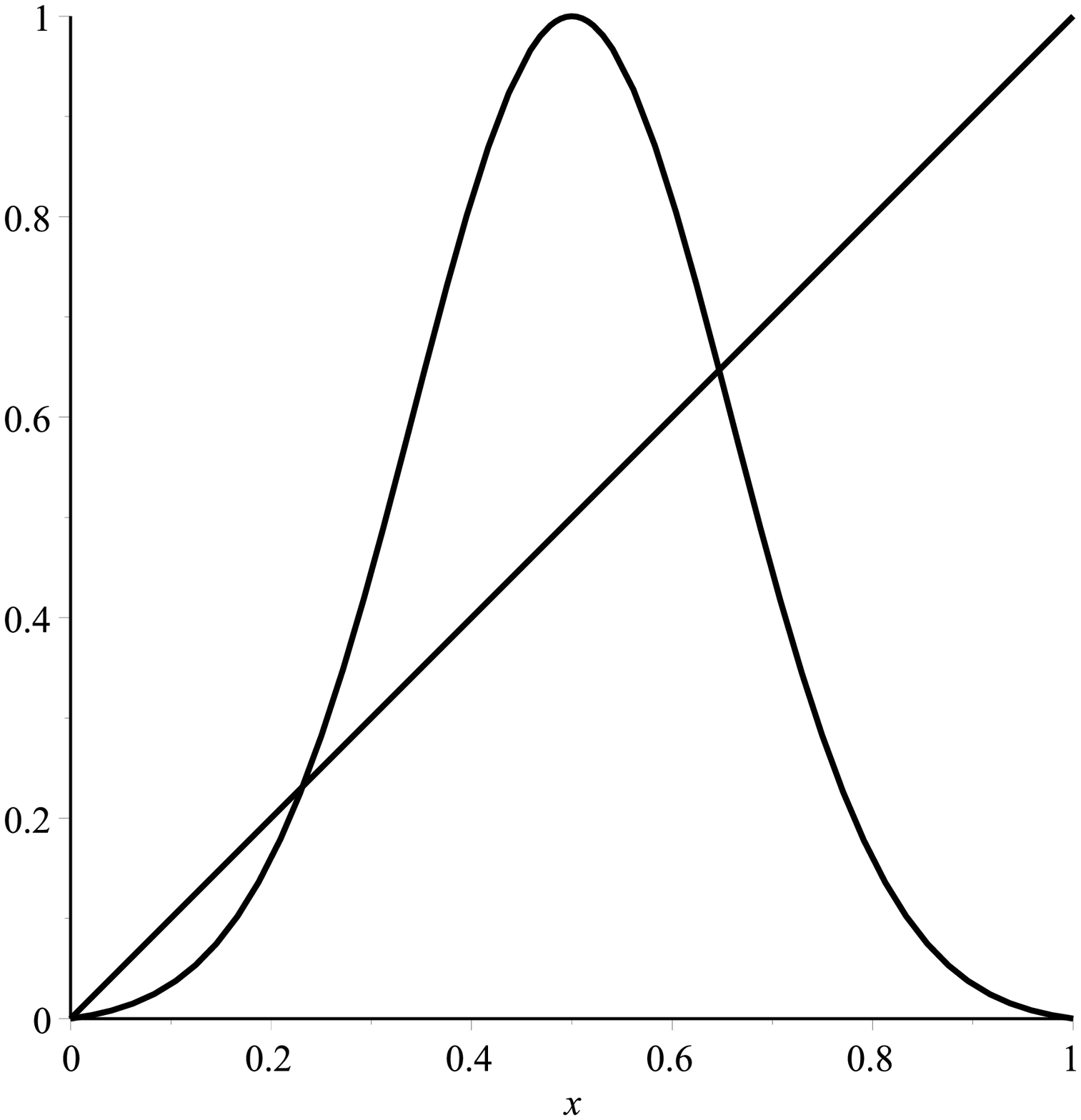}
\vspace{-3cm}
}
\centerline{Fig. 29: Ally effect}

The equation \eqref{ally} describes the dynamics of vanishing population, i.e. for any initial value $x_0\in(0,1)$ that is different from the equilibrium or cycles of the system \eqref{ally}, we have $x_n\to 0, n\to\infty.$

\centerline{
\includegraphics[scale=0.25]{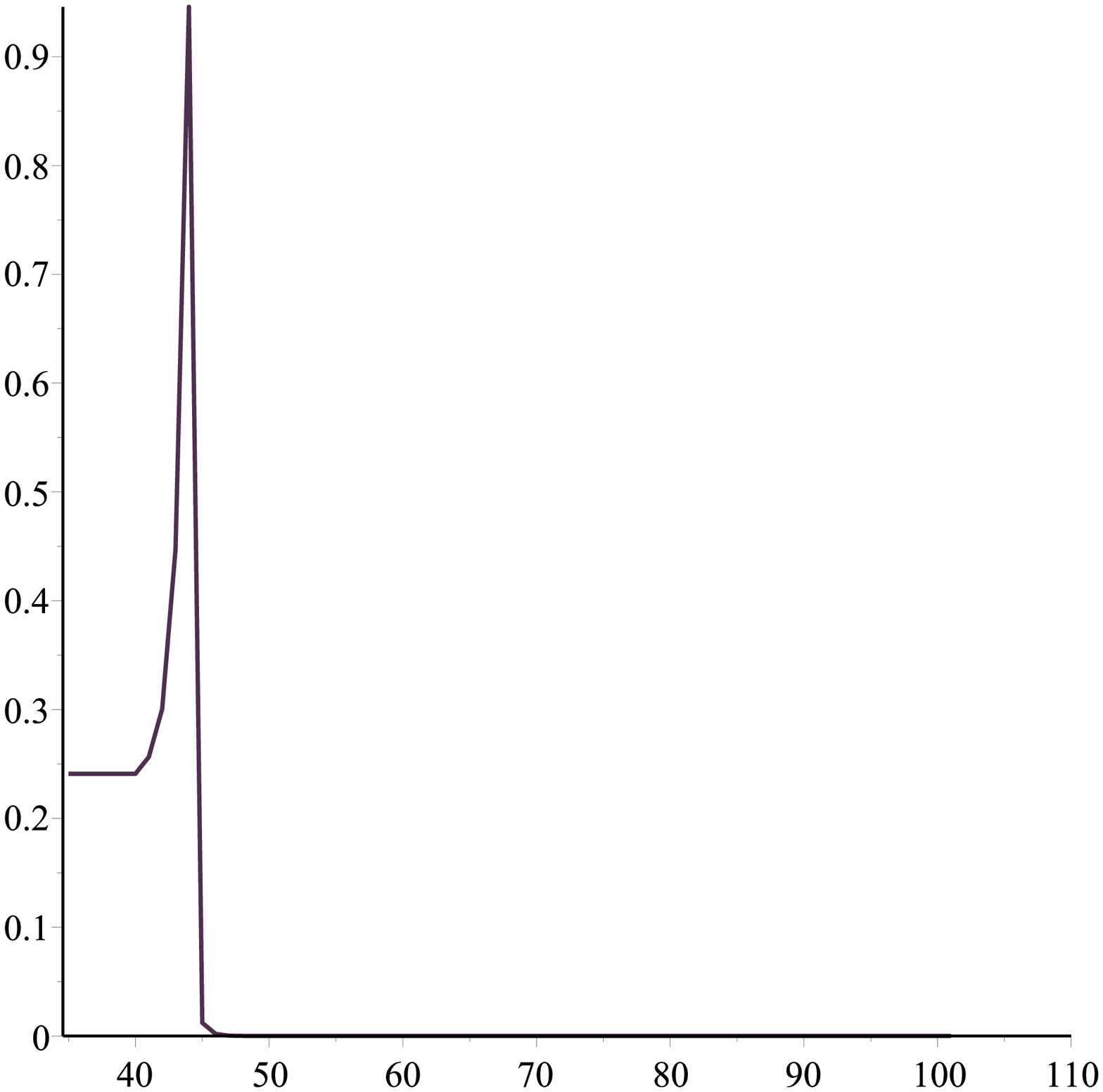}
%\vspace{-3cm}
}
\centerline{Fig. 30: Ally effect modeling, $N=2.$}
\bigskip

It is not difficult to see that for equilibrium $x^*\approx 0.647$ and the multiplier being negative we have $\mu^*\approx 3.84$ while the second multiplier is positive. To stabilize the equilibrium in the system \eqref{2} one should choose $N\ge 3.$ For $N=3$ one can find that {\it standard} coefficients $a_1^{(3)}\approx0.439,$ $a_2^{(3)}\approx0.414,$ $a_3^{(3)}\approx0.146.$ The rate of convergence is determined by the largest in absolute value of the root of the equation
$z^3+\mu^*(a_1^{(3)}z^2 + a_2^{(3)}z+a_3^{(3)})=0.$ This value is approximately $0.969.$ As we can see it is close to 1, therefore the convergence will be quite slow.\\

To increase the rate of convergence let us apply an algorithm from the previous section. We start with the root of the equation
$$
(a_1^{(3)}\rho + a_2^{(3)}\rho^2+a_3^{(3)}\rho^3)\cot^2\frac\pi8=\mu^*.
$$
It is $\rho_1\approx 0.766.$ Let us find new {\it modified} control coefficients
$$
b_j^{(3)}=\frac{a_j^{(3)}\rho_1^j}{a_1^{(3)}\rho_1 + a_2^{(3)}\rho_1^2+a_3^{(3)}\rho_1^3}: \;
b_1^{(3)}\approx 0.516, b_2^{(3)}\approx 0.379, b_3^{(3)}\approx 0.105.
$$
To determine the rate of convergence let us find the maximal in absolute value root of the equation
$$
z^3+\mu^*(b_1^{(3)}z^2 + b_2^{(3)}z+b_3^{(3)})=0.
$$
It is 0.761.\\

It is possible to increase a rate of convergence even more if we take $N=4.$ Then the {\it standard} coefficients are $a_1^{(4)}\approx 0.306,$ $a_2^{(4)}\approx 0.371,$
$a_3^{(4)}\approx 0.247,$ $a_4^{(4)}\approx 0.076.$ Let us find the root of the equation
$$
(a_1^{(4)}\rho + a_2^{(4)}\rho^2+a_3^{(4)}\rho^3 +a_4^{(4)}\rho^4)\cot^2\frac\pi{10}=\mu^*.
$$
It is $\rho_2\approx 0.623.$ Then
$$
b_j^{(4)}=\frac{a_j^{(4)}\rho_2^j}{a_1^{(4)}\rho_2 + a_2^{(4)}\rho_2^2+a_3^{(4)}\rho_2^3 +a_4^{(4)}\rho_2^4 }
$$
and
$$
b_1^{(4)}\approx 0.466, b_2^{(4)}\approx 0.356, b_3^{(4)}\approx 0.149, b_4^{(4)}\approx 0.029.
$$
The rate of convergence is about 0.618 as a maximal in absolute value root of the equation
$$
z^4+\mu^*(b_1^{(4)}z^3 + b_2^{(4)}z^2+b_3^{(4)}z+b_4^{(4)})=0.
$$

Solutions of the equation \eqref{2} with the choice of the controls,
i.e. corresponding to the system
$$
x_{n+1}=f\left(\sum_{j=1}^3 a_j^{(3)}x_{n-j+1}\right),\quad x_{n+1}=f\left(\sum_{j=1}^3 b_j^{(3)}x_{n-j+1}\right),\quad x_{n+1}=f\left(\sum_{j=1}^4 b_j^{(4)}x_{n-j+1}\right)
$$
are displayed on the figure 31 in pink, blue and red colors respectively.
\bigskip

\centerline{
\includegraphics[scale=0.25]{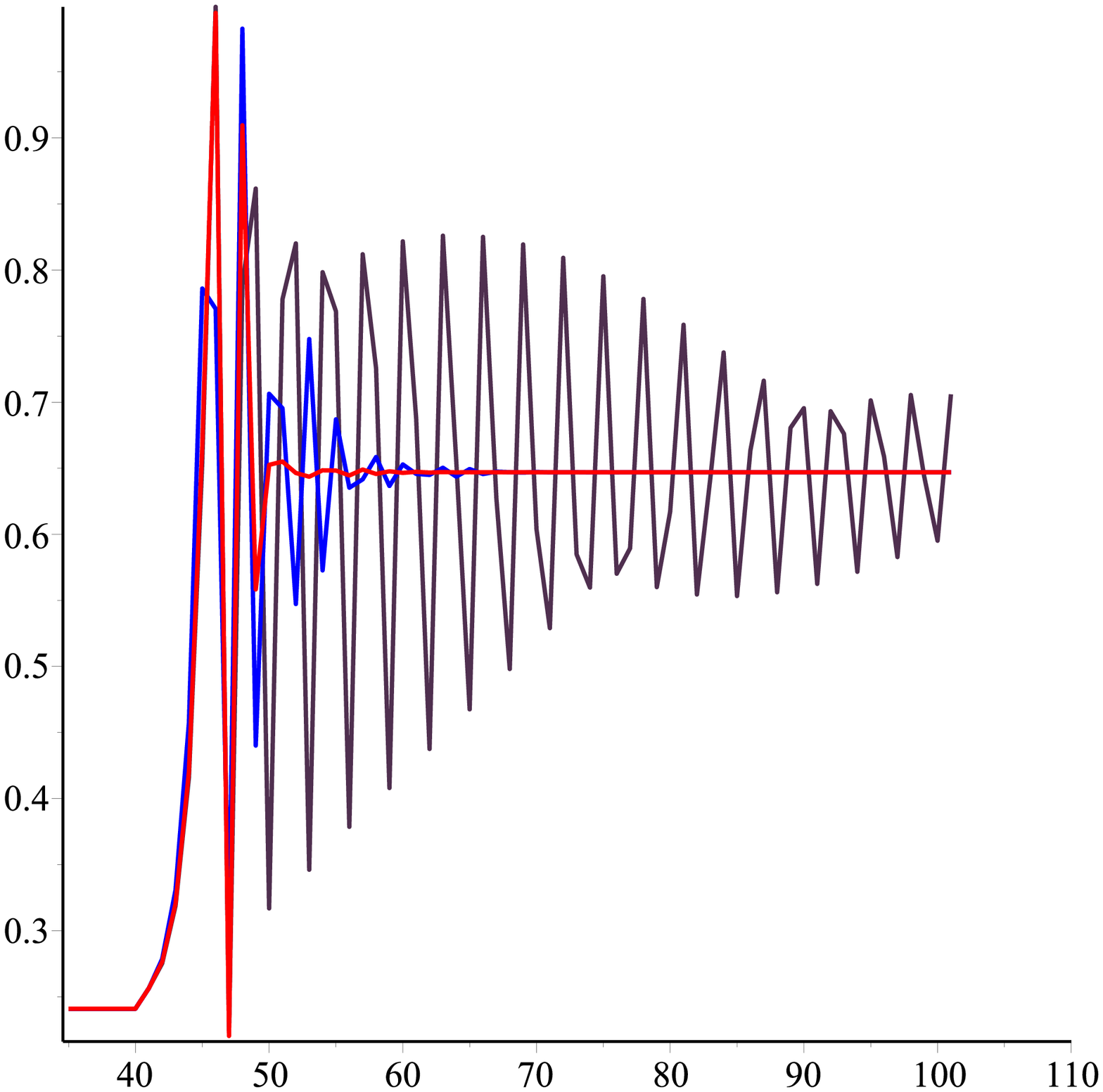}
%\vspace{-3cm}
}
\centerline{Fig. 31: Controlled system with different $\rho$.}

The difference is evident.

\subsection{Example 3}
The figures below demonstrates the difference in the rate of detecting of 3-cycles in the {\it standard} logistic equation.

\centerline{
\includegraphics[scale=0.25]{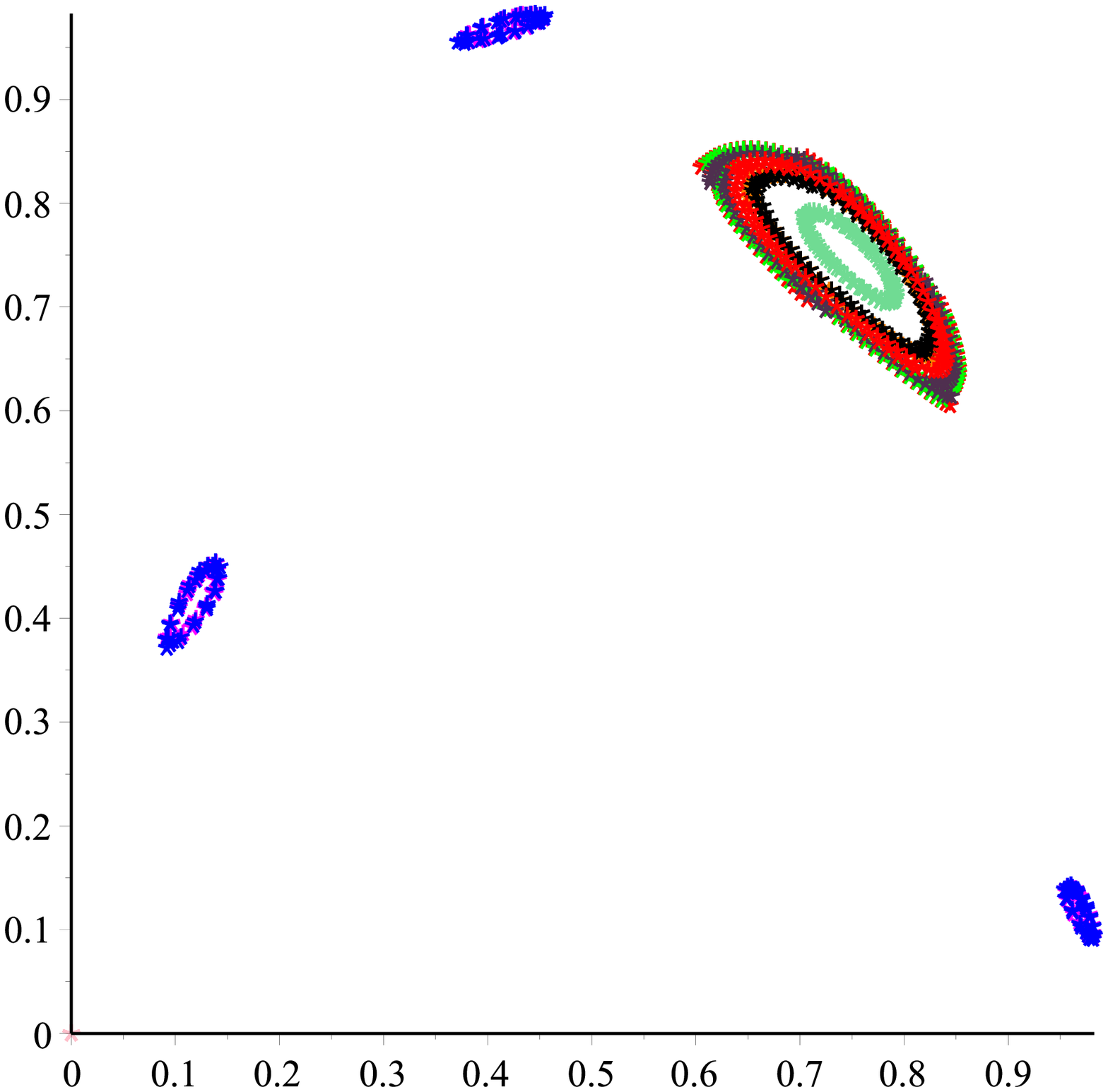}
%\vspace{-3cm}
}
\centerline{Fig. 32: Logistic equation: 3-cycle and equilibrium, {\it standard} control for $n=500,...,800$ }

\centerline{
\includegraphics[scale=0.25]{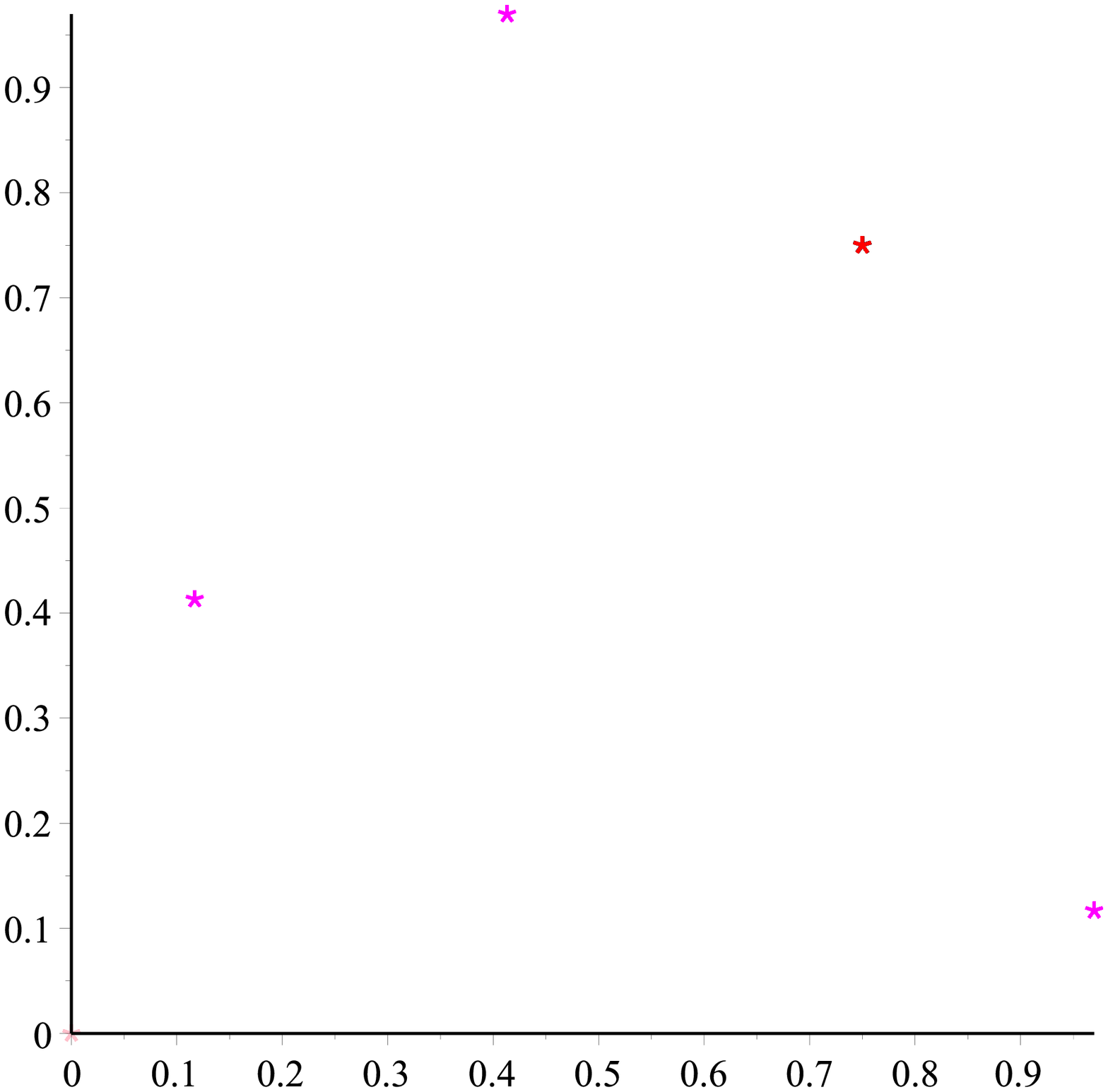}
%\vspace{-3cm}
}
\centerline{Fig. 33: Logistic equation: 3-cycle and equilibrium, $r=0.9$ control for $n=500,...,800$ }
\bigskip

\subsection{Example 4} Let us consider the equation of a sudden occurence  of chaos (SOC) \cite{TD}
$$
x_{n+1}=F(x_n),\quad F(x)=(1+\sqrt 2)\left(\frac12-\left|x-\frac12\right|\right)+x.
$$
To define the 2-cycle let us apply the system \eqref{2} with $N=4$ and $T=2.$ The mixing coefficients are computed for $\rho=1$ and $\rho=0.8$
$$
\left\{ 0.4375, 0.3125, 0.1875, 0.0625 \right\},\quad
\left\{ 0.5211, 0.2978, 0.1429, 0.0381 \right\}
$$
The dynamics is displayed on the figures 34 and 35. Its clearly visible how faster the control works for $\rho=0.8.$

\centerline{
\includegraphics[scale=0.25]{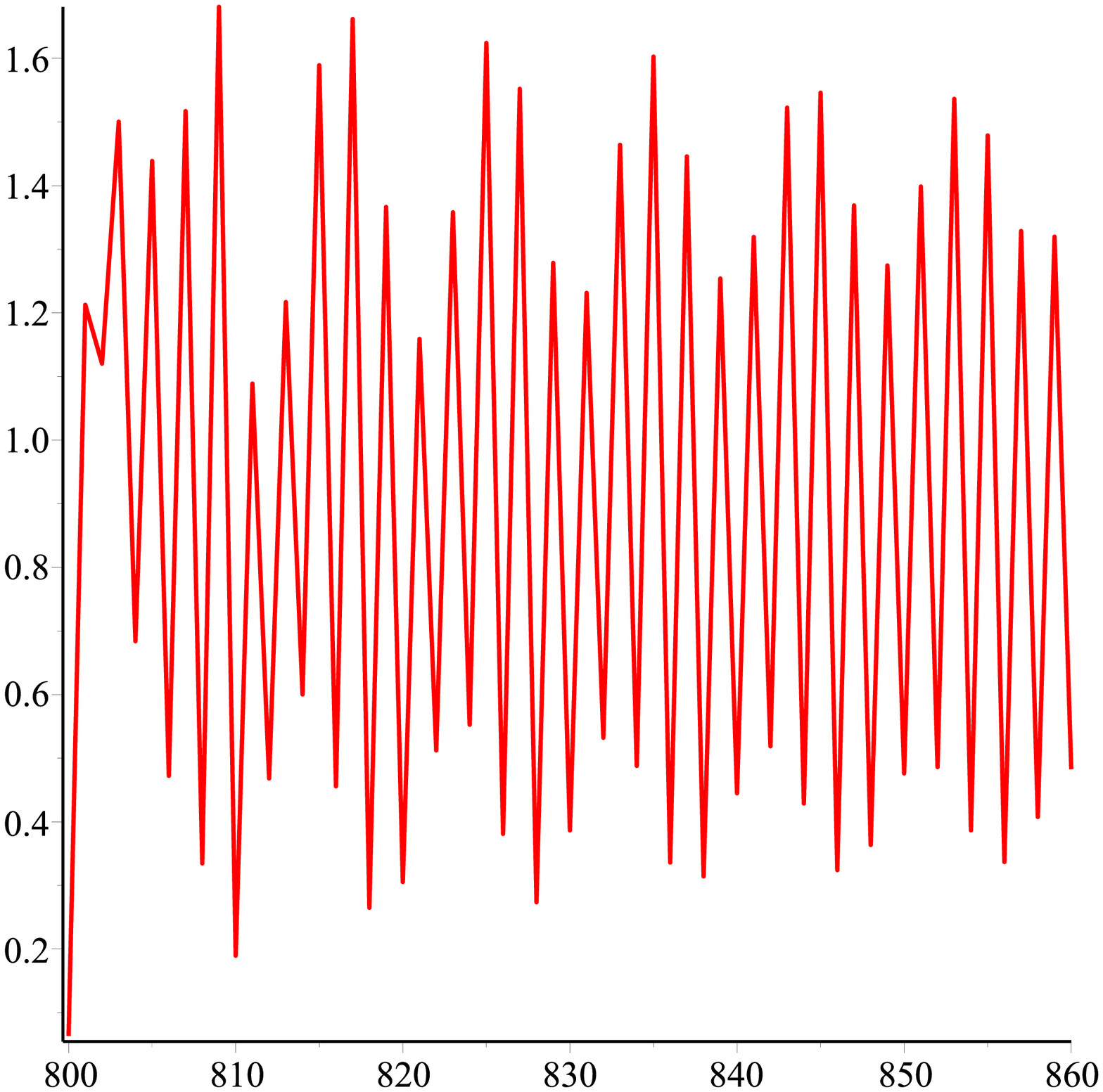}\hspace{1cm}
\includegraphics[scale=0.25]{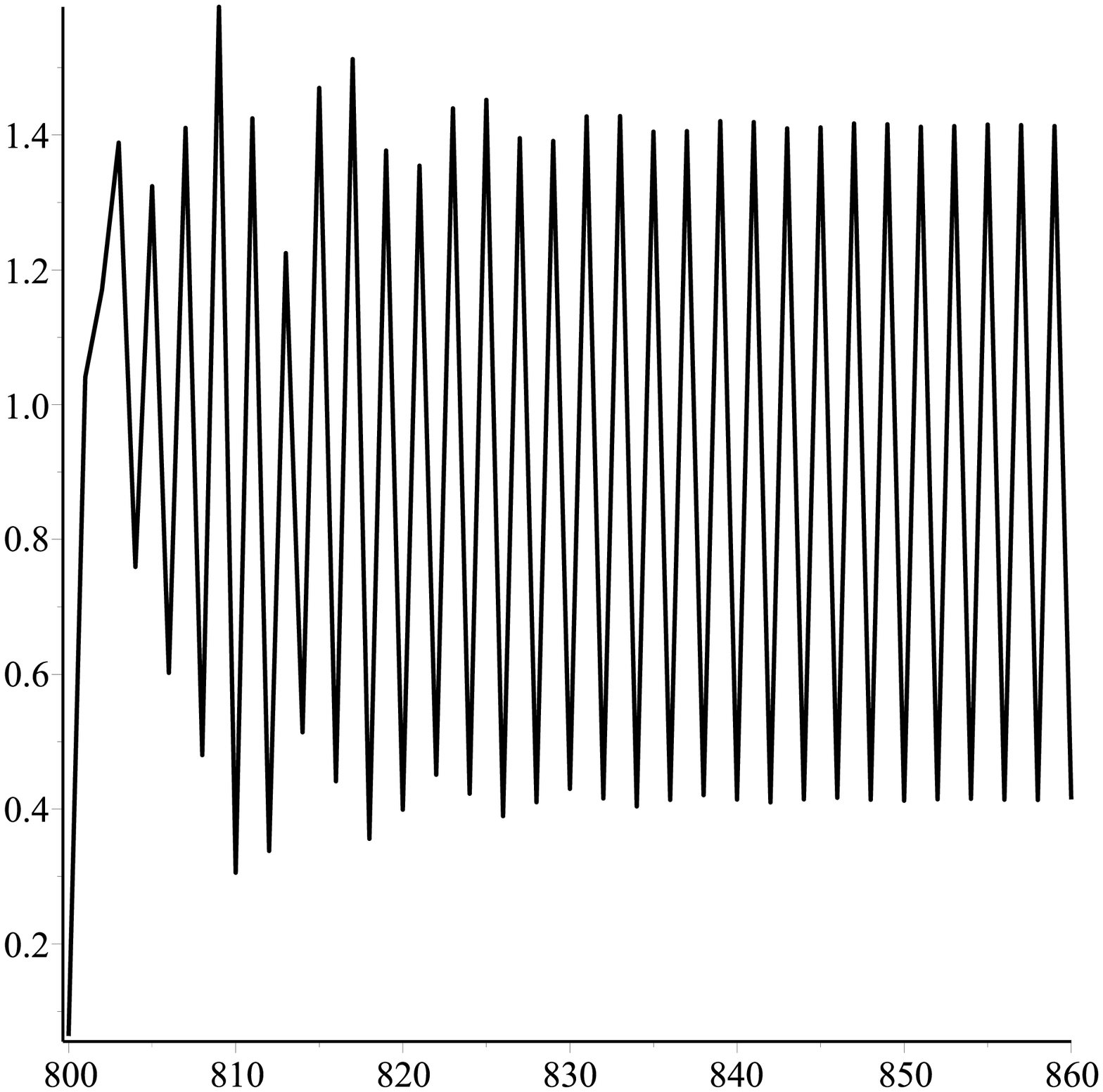}
%\vspace{-3cm}
}
\centerline{Fig. 34: SOC  $T=2, N=4$\hspace{1cm} Fig. 35: SOC  $T=2, N=4, \rho=0.8$  }
\bigskip

\section{Conditions of applying of the method}
The standard method of stabilization in theory should stabilize a cycle of any length $T$ for any value of the cycle multiplier. This follows from asymptotic estimates of the quantities $I_N^{(N)}$ in real and complex cases \cite{DKST}

$$
|I_N^{(T)}|\sim \frac1{N^2}\quad\mbox{ for large $N$ and $T,$}\; \{\mu_1,\ldots,\mu_m\}\in\{\mu\in\mathbb R:\mu\in(-\mu^*,0)\}
$$
$$
|I_N^{(T)}|\sim \frac1{N}\quad\mbox{ for large $N$ and $T,$}\; \{\mu_1,\ldots,\mu_m\}\in\{\mu\in\mathbb C:|\mu+R|<R\}
$$
If we want not just to stabilize the cycle but additionally guarantee the proper rate of convergence, then the quantities $\mu^*$ and $R$ characterize the diameter of the region of multipliers localizations, which cannot be made arbitrarily large regardless of the choice of $N.$ \\

Let us first demonstrate it for $T=1$ with real multipliers. Let us estimate the right hand side in the inequality
$$
\mu^*< \rho\sum_{k=1}^N a_k^{(N)}\rho^{k-1}\cdot\frac1{|I_N^{(1)}|},
$$
where $|I_N^{(1)}|=\tan^2\frac\pi{2(N+1)}$ and $a_j^{(N)}$ satisfies \eqref{aj1}. We have
$$
                \sum_{k=1}^N a_k^{(N)}\rho^k\cdot\frac1{|I_N^{(1)}|}< 2\cot\frac\pi{2(N+1)}\sum_{j=1}^N \frac{\pi j}{N+1} \rho^j=\frac{2\pi}{N+1}\cot\frac\pi{2(N+1)} \sum_{j=1}^N j\rho^j=
$$
$$
\frac{2\pi}{N+1}\cot\frac\pi{2(N+1)} \frac\rho{(1-\rho)^2}\cdot
\left(1-\rho^{N+1}-(N+1)\rho^N+(N+1)\rho^{N+1} \right).
$$
Therefore for the given radius $\rho$ the value $\mu^*$ characterizing the size of admissible set of multipliers location does not exit $\frac{4\rho}{(1-\rho)^2}$ regardless of $N.$ \\

We post the dependence of the value $\mu^*$ for the radii $\rho\in\{\frac12,\frac23,\frac9{10},1\}$ and $N=1,...,10$ in  Table 1 at the end of the article.
\bigskip

Let us consider the case $T=2, \mu$--real. In that case the inequality for the diameter of the sets of localization of multipliers has the form
$$
\mu^*< \rho\left(\sum_{k=1}^N a_k^{(N)}\rho^{k-1}\right)^2\cdot\frac1{|I_N^{(2)}|},
$$
where $|I_N^{(2)}|=\frac1{N^2}$ and $a_j^{(N)}$ are defined by \eqref{aj2}.\\

The sum $\sum_{k=1}^N a_k^{(N)}\rho^k$ can be computed explicitly
$$
\sum_{k=1}^N a_k^{(N)}\rho^k =\left(\frac2N+\frac1{N^2}\right)
\frac{\rho(1-\rho^N)}{1-\rho}-\frac2{N^2}\frac\rho{(1-\rho)^2}\cdot
$$
$$
\left(1-\rho^{N+1}-(N+1)\rho^N+(N+1)\rho^{N+1} \right).
$$
Then when $N\to\infty$
$$
\mu^*<\frac{N^2}\rho\left( \sum_{k=1}^N a_k^{(N)}\rho^k\right)^2\sim\frac{4\rho}
{(1-\rho)^2}.
$$
The value $\mu^*$ also is bounded for all $N$ if $\rho<1.$
We post the dependence of the value $\mu^*$ for the radii $\rho\in\{\frac12,\frac23,\frac9{10},1\}$ and $N=1,...,10$ in Table 2 at the end of the article.
\bigskip

\iffalse%%%%%%%%%%%%%%%%%%%%%%%%%%
The complex case is considered in the same way. For $T=1$
$$
\mu^*<\rho \sum_{k=1}^N a_k^{(N)}\rho^{k-1}\cdot\frac1{2|I_N^{(1)}|},
$$
where $|I_N^{(1)}|=\frac1N$ and $a_k^{(N)}=\frac2N\left(1-\frac k{N+1}\right), k=1,...,N.$\\

Here is the procedure. First, for given $\rho$ the value of $N$ is defined by the condition
$$\mu^*_{N-1}\le\mu<\mu^*_N.$$
Then, after determining the {\it standard} coefficients $a_j^{(N)}$ by formulas \eqref{aj1} or similar formulas we determine the mixing coefficients
$$
b_j=\frac1{\sum_{k=1}^N a_k^{(N)}\rho^k} a_j^{(N)}\rho^j,\quad j=1,...,N.
$$
Note that
$$
|a_j^{(N)}|\le \frac{\pi^2}{\sqrt 2}\frac{j}{N^2},
$$
therefore
$$
\mu^*< \frac{\pi^2}{\sqrt 2 N^2}\cot^2\frac\pi{2(N+1)}\sum_{j=1}^N j\rho^j.
$$
Thus, $\mu^*$ is bounded by the constant that depends on $\rho$ and is independent on $N$ if $\rho<1$, in contrast to the case $\rho=1.$\\
\fi%%%%%%%%%%%%%%%%%%%

The complex case is considered in the same way. For $T=1$
$$
R<\rho\sum_{k=1}^N a_k^{(N)}\rho^{k-1}\cdot \frac1{2|I_N^{(1)}|}
$$
where $|I_N^{(1)}|=\frac1N$ and $a_k^{(N)}$ are defined by \eqref{aj3}.

Let us compute
$$
\sum_{k=1}^N a_k^{(N)}\rho^k =\frac2N
\frac{\rho(1-\rho^N)}{1-\rho}-\frac2{N(N+1)}\frac\rho{(1-\rho)^2}\cdot
$$
$$
\left(1-\rho^{N+1}-(N+1)\rho^N+(N+1)\rho^{N+1} \right).
$$
From there
$$
R<\frac{\rho(1-\rho^N)}{1-\rho}-\frac1{N+1}\frac\rho{(1-\rho)^2}
\left(1-\rho^{N+1}-(N+1)\rho^N+(N+1)\rho^{N+1} \right).
$$
Asymptotically, when $N\to\infty$
$$
R<\frac\rho{1-\rho}.
$$
Thus, in the case of complex multipliers the diameter of the region of  multiplier locations is bounded for $\rho<1$ by a value independent of $N.$ The table of $R$ values for different $\rho$ and $N$ are at the end of the article.\\

\bigskip

Thus, it is shown that dramatical improvement of the rate of convergence is possible only for relatively small regions of multipliers. Conversely, if the region of a multiplier's location is large enough then to place the roots of the characteristic polynomials in a disc of small radius is impossible for any $N.$

\section{Conclusion}
In this paper we consider a generalization of non-linear delay feedback control developed in \cite{DK, DHKS, DKST, DSS}. We show that one can modify the coefficients of the mixing or delay feedback control to increase the rate of convergence to $T$-cycles of  interest. Also, we found the range of limitation of the modified control. The price we pay for the acceleration of convergence is an increase of the depth of necessary prehistory.

\section{Acknowledgement}

The authors would like to thank Paul Hagelstein for his interest in the subject and for his help in preparation of this manuscript.

\newpage

\centerline{\small
\begin{tabular}{|p{0.75in}||p{0.75in}|p{0.75in}|p{0.75in}|p{0.75in}|} \hline
&&&&\\
$N$ & $\mu^*_N(1)$ & $\mu^*_N(0.9)$ & $\mu^*_N(2/3)$&  $\mu^*_N(1/2)$ \\
&&&&\\
\hline\hline
&&&&\\
1 &1.& .9& .6666666667& .5000000000\\
&&&&\\
\hline
&&&&\\
2&3.000000000& 2.610000000& 1.777777778& 1.250000000\\
&&&&\\
\hline
&&&&\\
3 &5.828427123& 4.882347562& 3.032995295& 1.990577650\\
&&&&\\
\hline
&&&&\\
4 &9.472135954& 7.631892583& 4.327466156& 2.663259414\\
&&&&\\
\hline
&&&&\\
5 &13.92820323& 10.79130682& 5.601974199& 3.252984336\\
&&&&\\
\hline
&&&&\\
6 &19.19566935& 14.30194674& 6.822475375& 3.761456200\\
&&&&\\
\hline
&&&&\\
7 &25.27414236& 18.11176689& 7.970659385& 4.196794604\\
&&&&\\
\hline
&&&&\\
8 &32.16343748& 22.17436353& 9.038213110& 4.568918033\\
&&&&\\
\hline
&&&&\\
9 &39.86345818& 26.44832483& 10.02310080& 4.887514234\\
&&&&\\
\hline
&&&&\\
10 &48.37415005& 30.89670701& 10.92711594& 5.161240519\\
&&&&\\
\hline
\end{tabular}
}
\bigskip

\centerline{Table 1 of critical values, T=1, $\mu\in(-\mu^*,0)$}

\newpage

\centerline{\small
\begin{tabular}{|p{0.75in}||p{0.75in}|p{0.75in}|p{0.75in}|p{0.75in}|} \hline
&&&&\\
$N$ & $\mu^*_N(1)$ & $\mu^*_N(0.9)$ & $\mu^*_N(2/3)$&  $\mu^*_N(1/2)$ \\
&&&&\\
\hline\hline
&&&&\\
1 & 1.& .9& .6666666667& .5000000000\\
&&&&\\
\hline
&&&&\\
2&4.000000000& 3.422250000& 2.240740741& 1.531250000\\
&&&&\\
\hline
&&&&\\
3 &9.000000000& 7.242009999& 4.105166897& 2.531250000\\
&&&&\\
\hline
&&&&\\
4 &16.00000000& 12.08735331& 5.963020122& 3.363769533\\
&&&&\\
\hline
&&&&\\
5 &25.00000000& 17.73121275& 7.684296098& 4.025703125\\
&&&&\\
\hline
&&&&\\
6 &36.00000000& 23.98439167& 9.223014722& 4.546997075\\
&&&&\\
\hline
&&&&\\
7 &49.00000000& 30.68965059& 10.57385227& 4.959843852\\
&&&&\\
\hline
&&&&\\
8 &64.00000000& 37.71670341& 11.74969591& 5.290775772\\
&&&&\\
\hline
&&&&\\
9 &81.00000000& 44.95798306& 12.77026032& 5.559896679\\
&&&&\\
\hline
&&&&\\
10 &100.0000000& 52.32505773& 13.65652485& 5.781992354\\
&&&&\\
\hline
\end{tabular}
}
\bigskip

\centerline{Table 2 of critical values, T=2, $\mu\in(-\mu^*,0)$}
\newpage

\centerline{\small
\begin{tabular}{|p{0.75in}||p{0.75in}|p{0.75in}|p{0.75in}|p{0.75in}|} \hline
&&&&\\
$N$ & $R(1)$ & $R(0.9)$ & $R(2/3)$&  $R(1/2)$ \\&&&&\\
\hline\hline&&&&\\
1&.5000000000& .4500000000& .3333333334& .2500000000\\&&&&\\
\hline&&&&\\
2&1.000000000& .8700000000& .5925925925& .4166666667\\&&&&\\
\hline&&&&\\
3&1.500000000& 1.262250000& .7962962965& .5312500000\\&&&&\\
\hline&&&&\\
4&2.000000000& 1.628820000& .9580246915& .6125000000\\&&&&\\
\hline&&&&\\
5&2.500000000& 1.971615000& 1.087791495& .6718750000\\&&&&\\
\hline&&&&\\
6&3.000000000& 2.292388715& 1.193023712& .7165178570\\&&&&\\
\hline&&&&\\
7&3.500000000& 2.592756112& 1.279263832& .7509765625\\&&&&\\
\hline&&&&\\
8&4.000000000& 2.874204890& 1.350674864& .7782118050\\&&&&\\
\hline&&&&\\
9&4.500000000& 3.138105961& 1.410404918& .8001953120\\&&&&\\
\hline&&&&\\
10&5.000000000& 3.385723059& 1.460851464& .8182705965\\&&&&\\
\hline
\end{tabular}
}
\bigskip

\centerline{Table 3 of critical values, T=1, $R>0.$}

\bigskip

D. Dmitrishin and E. Franzheva, Odessa Polytechnic University, Odessa, 65044, Ukraine. E-mail: dmitrishin@opu.ua\\

%P.Haglestein, Baylor Uinversity, Waco, TX, paul${}_-$hagelstein@baylor.edu \\

A. Stokolos, Georgia Southern University, Statesboro, GA, 30460. E-mail:  astokolos@georgiasouthern.edu\\

\end{document}